\documentclass[twocolumn,showpacs,amsmath,amssymb,prx,footinbib,floatfix,superscriptaddress]{revtex4-1}
\usepackage{graphicx}
\usepackage{braket}
\usepackage{savesym}
\usepackage{amsfonts}
\usepackage{bm}
\usepackage{color}
\usepackage{subfigure}
\usepackage{wasysym}
\usepackage{upgreek}
\usepackage{float}
\usepackage{dsfont}
\usepackage{mathtools}% Loads amsmath
\usepackage{multirow}
\usepackage[usenames,dvipsnames,svgnames,table]{xcolor}
\usepackage{hyperref}
\usepackage{hhline} %for double horizontal lines without vertical break in tabular environment
\usepackage{multirow} %for cells that span multiple rows in columns

\newcommand{\Eq}[1]{Eq.~(\ref{#1})}
\newcommand{\Fig}[1]{Fig.~\ref{#1}}
\newcommand{\tr}[0]{\textrm{Tr}}

\usepackage{makecell}

\usepackage[english]{babel}

\hypersetup{
	colorlinks=true,       % false: boxed links; true: colored links
	linkcolor=VioletRed,  % color of internal links (change box color with linkbordercolor)
    citecolor=JungleGreen,        % color of links to bibliography
    filecolor=Orchid,      % color of file links
    urlcolor=black           % color of external links
}
\begin{document}

\title{Recurrent Neural Network Wave Functions} 

\author{Mohamed Hibat-Allah}
\email{mohamed.hibat.allah@uwaterloo.ca}

\affiliation{Vector Institute, MaRS Centre, Toronto, Ontario, M5G 1M1, Canada}
\affiliation{Perimeter Institute for Theoretical Physics, 31 Caroline Street North, Waterloo, Ontario, N2L 2Y5, Canada} 
\affiliation{Department of Physics and Astronomy, University of Waterloo, Ontario, N2L 3G1, Canada}

\author{Martin Ganahl}
\affiliation{Perimeter Institute for Theoretical Physics, 31 Caroline Street North, Waterloo, Ontario, N2L 2Y5, Canada}

\author{Lauren E. Hayward}
\affiliation{Perimeter Institute for Theoretical Physics, 31 Caroline Street North, Waterloo, Ontario, N2L 2Y5, Canada}

\author{Roger G. Melko}
\affiliation{Perimeter Institute for Theoretical Physics, 31 Caroline Street North, Waterloo, Ontario, N2L 2Y5, Canada}
\affiliation{Department of Physics and Astronomy, University of Waterloo, Ontario, N2L 3G1, Canada}

\author{Juan Carrasquilla}
\affiliation{Vector Institute, MaRS Centre, Toronto, Ontario, M5G 1M1, Canada}
\affiliation{Department of Physics and Astronomy, University of Waterloo, Ontario, N2L 3G1, Canada}

\date{\today}

%%%%%%%%%%%%%%%%%% ABSTRACT %%%%%%%%%%%%%%%%%%
\begin{abstract}

A core technology that has emerged from the artificial intelligence revolution is the recurrent neural network (RNN).
Its unique sequence-based architecture provides a tractable likelihood estimate with stable training paradigms, 
a combination that has precipitated many spectacular advances in natural language processing and neural machine translation.
This architecture also makes a good candidate for a variational wave function, where the RNN parameters are tuned to learn the approximate ground state of a quantum Hamiltonian. In this paper, we demonstrate the ability of RNNs to represent several many-body wave functions, optimizing the variational parameters using a stochastic approach.  Among other attractive features of these variational wave functions, their autoregressive nature allows for the efficient calculation of physical estimators by providing independent samples. We demonstrate the effectiveness of RNN wave functions by calculating ground state energies, correlation functions, and entanglement entropies for several quantum spin models of interest to condensed matter physicists in one and two spatial dimensions.

\end{abstract}

\maketitle

%%%%%%%%%%%%%%%%%% INTRODUCTION %%%%%%%%%%%%%%%%%%
\section{Introduction}

The last decade has marked the start of a worldwide artificial intelligence (AI) revolution, which is dramatically affecting industry, science, and society.
The source of the current AI resurgence can largely be traced back to AlexNet~\cite{hinton2012}, one of the most influential breakthrough papers in computer vision, which provided a dramatic quantitative improvement in object recognition tasks and
popularized the paradigm of deep learning~\cite{LeCun2015}.
The concept of deep learning encompasses a set of machine learning techniques where data are processed through the composition of parametrized nonlinear layers, each of which generates increasingly abstract representations of the original data~\cite{LeCun2015}.
This paradigm has demonstrated an unprecedented unifying power by making advances in areas as diverse as image recognition~\cite{he2016}, natural language processing~\cite{young2018}, drug discovery~\cite{vamathevan2019}, self-driving cars~\cite{selfdrivingcar2019}, game play~\cite{silver2017mastering}, and more.

The striking performance of deep learning methods has motivated researchers to use a machine learning perspective to reexamine problems in the physical sciences, including areas such as particle physics, cosmology, materials science, quantum chemistry, and statistical physics~\cite{carleo2019machine}. The exploration of machine learning techniques has been particularly prominent in the field of quantum many-body physics, where the task of elucidating the equilibrium and non-equilibrium properties of interacting many-particle systems remains at the research frontier of quantum information and condensed matter physics.
One of the first successful technology transfers from machine learning into many-body physics involved the use of neural network methods in 
a variational calculation~\cite{carleo2017}.
The variational principle is the theoretical bedrock behind many of the most powerful numerical approaches to solving many-body problems in quantum mechanics~\cite{PhysRev.136.B864,PhysRev.108.1175,PhysRevLett.50.1395}. 
Modern incarnations range from well-established techniques such as variational Monte Carlo (VMC)~\cite{becca2017} and tensor networks (TN)~\cite{orus2019} to variational quantum eigensolvers (VQE) for quantum computation~\cite{peruzzo2014a}.  
The resurgence of interest in machine learning has motivated a rich new playground for variational calculations based on neural networks~\cite{androsiuk1993,LAGARIS19971,sugawara2001,carleo2017,melko2019}.
Simultaneous to the computer vision revolution, a wide array of model architectures and algorithmic advances have also emerged in the context of natural language processing (NLP) -- the technology that enables computers to process and understand human language. Some of the most important algorithmic advances in NLP have been developed in the context of sequence learning using {\it recurrent neural networks} (RNNs)~\cite{hochreiter1997long,graves2012supervised,cho2014learning,chung2014,lipton2015}.
These have resulted in impressive results in speech and text comprehension, as well as in state-of-the-art results in neural machine translation. 
With RNNs and other algorithmic and conceptual advances, algorithms are bringing machine translation and speech recognition closer to the human level with unprecedented success~\cite{chung2014,Transformer,devlin2018bert,yang2019xlnet}. Here we explore whether the power and scalability of NLP models such as the RNN can be extended to  applications in physical systems, in particular to perform variational calculations to find the low-energy states of quantum many-body Hamiltonians.

RNNs have already proven to be powerful tools within the field of many-body physics. In Ref.~[\onlinecite{Carrasquilla2019}], RNNs were applied in the context of quantum state tomography and were found to be capable of representing a broad range of complex quantum systems, including prototypical states in quantum information and ground states of local spin models.
Furthermore, RNNs have established similarities to matrix product states (MPS) and are capable of capturing entanglement properties of quantum many-body systems~\cite{PhysRevLett.122.065301}.
To date however, little effort has been made to develop NLP technology for use together with the variational principle.
Here we investigate the power of RNNs and their extensions for approximating the ground state of strongly correlated local Hamiltonians.
We demonstrate how the variational principle can be combined with RNNs to yield highly efficient ansatz wave functions. Our proposal makes use of the {\it autoregressive} property \cite{Bengio2000,NADE,Wu_2019} of RNNs, which, unlike traditional VMC methods, allows for sampling from the wave function. 
We variationally optimize our RNNs to approximate ground states of various strongly correlated quantum systems in one and two dimensions. 
We find excellent agreement for local correlation functions and entanglement entropy upon comparison with well-established state-of-the-art approaches, while requiring only a fraction of the variational parameters. 
Through extensive scaling studies, we show that the intrinsic bias of our ansatz can be systematically reduced to yield highly accurate ground state approximations of large quantum systems.

%%%%%%%%%%%%%%%%%% RNN %%%%%%%%%%%%%%%%%%

\section{Classical and quantum recurrent neural networks}
\subsection{RNNs for classical probability distributions}
\label{sec:RNN_wf}
We consider probability distributions defined over a discrete sample space, where a single configuration consists of a list $\bm{\sigma}\equiv (\sigma_1,\sigma_2,\dots, \sigma_N)$ of $N$ variables $\sigma_n$,
and $\sigma_n\in\{0,1,\dots ,d_v-1\}$. Here, the {\it input dimension} $d_v$ represents the number of possible values that 
any given variable $\sigma_n$ can take. 
A central task in machine learning is to use a set of empirical samples to infer probability distributions in cases where there are strong correlations among the variables $\sigma_n$.
We denote the probability of a configuration $\bm{\sigma}$ by 
$P(\bm{ \sigma})\equiv P(\sigma_1, \sigma_2,\dots ,\sigma_N)$, and use the product rule for probabilities to express this distribution as 
\begin{align}
    P(\bm{ \sigma})= P(\sigma_1)P(\sigma_2|\sigma_1) \cdots P(\sigma_N|\sigma_{N-1}, \dots, \sigma_2, \sigma_1),
    \label{eq:prod_rule}
\end{align}
where 
$P(\sigma_i|\sigma_{i-1}, \dots, \sigma_2, \sigma_1)\equiv P(\sigma_i|\sigma_{<i})$
is the conditional distribution of $\sigma_i$ given a configuration of all $\sigma_j$ with $j<i$.

Specifying every conditional probability $P(\sigma_i| \sigma_{<i})$ gives a full characterization of any possible distribution $P(\bm{ \sigma})$, but in general such a representation grows exponentially with system size $N$. 
Typically, real-world distributions are assumed to endow enough structure on the problem to allow for accurate 
approximate descriptions of $P(\bm{\sigma})$ that use far fewer resources~\cite{Goodfellow-et-al-2016}.
This assumption is also applicable in the context of ground state wave functions that arise in physical systems, which
we will discuss at length in this paper.

RNNs form a class of correlated probability distributions of the form \Eq{eq:prod_rule},
where the $P(\bm \sigma)$ are entirely specified through the conditionals $P(\sigma_i| \sigma_{<i})$. 
The elementary building block of an RNN is a {\it recurrent cell}, that has emerged in different versions in the past~\cite{lipton2015}. In its simplest form, a recurrent cell is a non-linear function
that maps the direct sum (or concatenation) of an incoming {\it hidden} vector $\bm{h}_{n-1}$ of dimension $d_h$ and an input vector $\bm{\sigma}_{n-1}$ to an output hidden vector $\bm{h}_{n}$ of dimension $d_h$ such that
\begin{equation}
    \bm{h}_{n} = f\left(W 
    [\bm{h}_{n-1} ; \bm{\sigma}_{n-1}] + \bm{b}\right),
    \label{eq:rnn_action}
\end{equation}
where $f$ is a non-linear {\it activation function}.

\begin{figure}[t]
    \centering
    \includegraphics[width=\linewidth]{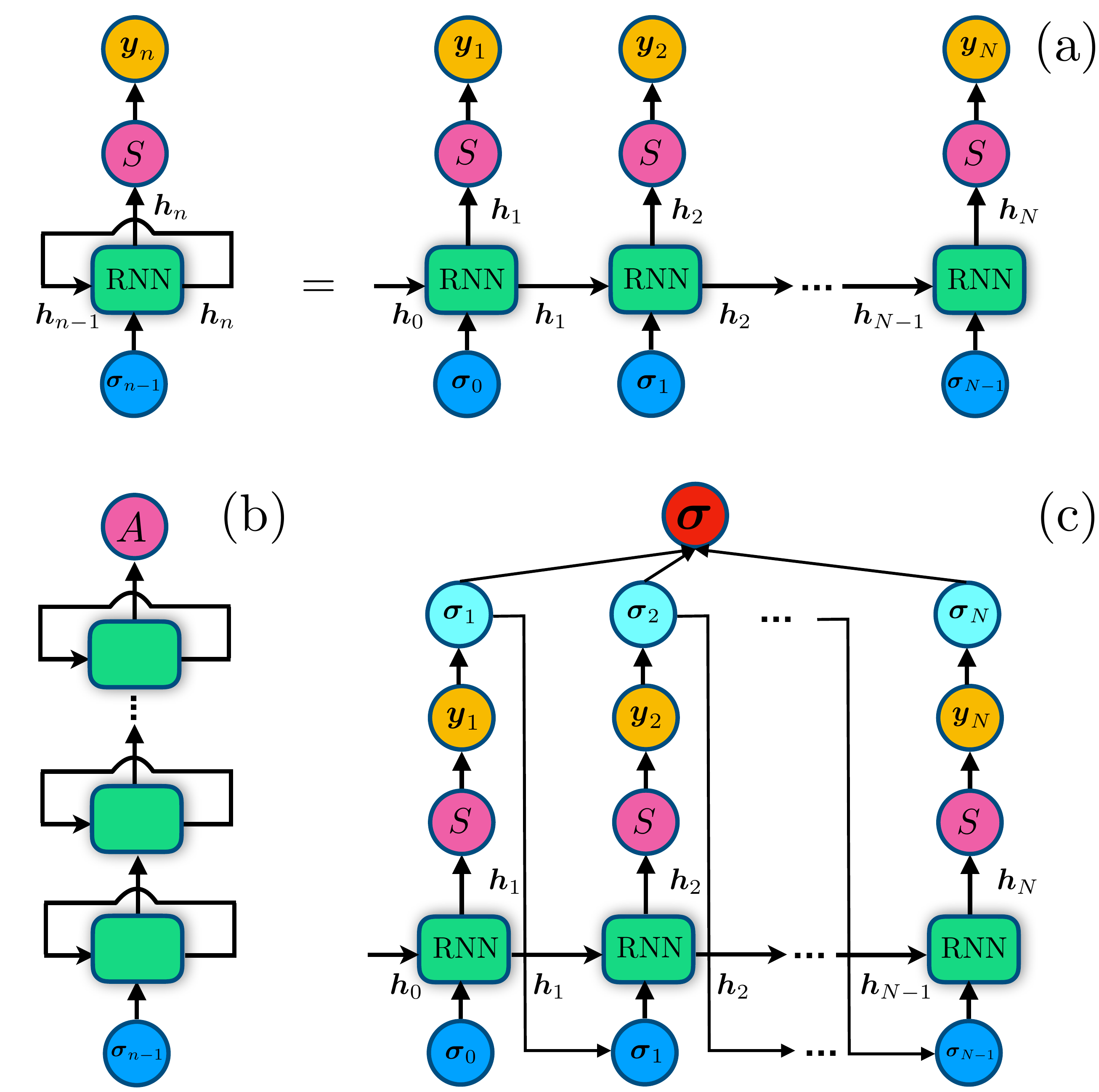}
    \caption{(a) Left-hand side: An RNN cell (green box) takes a sequence of inputs $\{\bm{\sigma}_n\}$, where at each step $n$ the input $\bm{\sigma}_{n-1}$ and the vector $\bm{h}_{n-1}$ are fed in the RNN cell which generates a vector $\bm{h}_n$ called the hidden state of the RNN. $\bm{h}_n$ is meant to encode the history of the previous inputs $\bm{\sigma}_{n'<n}$. Moreover, the hidden state $\bm{h}_n$ is fed to a fully connected layer with Softmax activation $S$ (magenta circles) to compute conditional probabilities. Right-hand side: The unrolled version of the RNN layer on the left-hand side. (b) A deep RNN model with $N_l$ stacked single RNN cells (green blocks) followed by a fully connected layer with activation function $A$ (magenta circle). Each single RNN cell at the $\ell$-th layer has its corresponding hidden state $\boldsymbol{h}^{\ell}_{n}$, which serves also as an input for the RNN cell at the $(\ell+1)$-th layer. (c) A graphical representation of autoregressive sampling of RNNs.}
    \label{fig:RNN}
\end{figure}

The parameters of this simple RNN (vanilla RNN) are given by the
weight matrix $W \in \mathbb{R}^{d_h \times (d_h + d_v)}$, the bias vector $\bm{b}\in \mathbb{R}^{d_h}$, and the states $\bm{h}_0$ and $\bm{\sigma}_0$ that initialize the recursion. In this paper, we fix $\bm{h}_0$ and $\bm{\sigma}_0$ to constant values. The vector $\bm{\sigma}_n$ is a one-hot encoding of the input $\sigma_n$ such that, e.g., 
$\bm{\sigma}_n = (1,0), (0,1)$ for $\sigma_n = 0,1$ (respectively) when the input dimension is two. 
The computation of the full probability $P(\bm \sigma)$ is carried out by sequentially computing
the conditionals, starting with $P(\sigma_1)$, as
\begin{equation*}
P\left(\sigma_n | \sigma_{n-1}, \dotsc , \sigma_{1} \right) = \bm{y}_{n} \cdot \bm{\sigma}_n,
\end{equation*}
where the right-hand side contains the usual scalar product between vectors and
\begin{equation}
    \bm{y}_{n} \equiv S\left( U\boldsymbol{h}_{n} + \bf{c}\right).\label{eq:softmax_layer}
\end{equation}
Here, $U\in \mathbb{R}^{d_v\times d_h}$ and $c\in \mathbb{R}^{d_v}$ are weights and biases of a so-called Softmax layer, and the Softmax activation function $S$ is given by \[
\text{S}(v_n) = \frac{\exp(v_n)}{\sum_i \exp(v_i)}.
\]
In Eq.~\eqref{eq:softmax_layer} $\bm y_n=(y_n^1, \dots, y_n^{d_v})$ is a $d_v$-component vector of  positive, real numbers summing up to $1$, i.e.,
\begin{align}
    \lVert\bm y_n\rVert_1 =1,
\end{align}
and thus forms a probability distribution over the states $\sigma_n$.
Once the vectors $\bm y_n$ have been specified, the full probability $P(\bm \sigma)$ is given by
\[
P(\bm{\sigma}) = \prod_{n=1}^N \bm{y}_n \cdot \bm{\sigma}_{n}.
\]
Note that $P(\bm \sigma)$ is already properly normalized to unity such that
\begin{align}
  \lVert P(\bm \sigma)\rVert_1 = 1.
\end{align}
Sampling from an RNN probability distribution is achieved in a similar sequential fashion.
To generate a sample $\bm \sigma= (\sigma_1,\dots ,\sigma_N)$ consisting of a set of $N$
configurations $\sigma_n$, one first calculates the hidden state $\bm{h}_1$
and the probability $\bm{y}_1$ from the initial vectors $\bm{h}_0$ and $\bm{\sigma}_0$.
 A sample $\sigma_1$ from the probability distribution  $\bm{y}_1$ is drawn, which is then fed as
 a one-hot vector $\bm{\sigma}_1$ along with $\bm{h}_1$ back into the recurrent cell to obtain $\bm{y}_2, \bm{h}_2$ and then $\sigma_2$. The procedure is then iterated until $N$ configurations $\sigma_n$ have been obtained as illustrated in Fig.~\ref{fig:RNN}(c).

From Eqs.~\eqref{eq:rnn_action} and \eqref{eq:softmax_layer}, it is evident that
the hidden vector $\bm{h}_{n}$ encodes information about previous spin configurations $\sigma_{<n}$.
For correlated probabilities, the history  $\sigma_{<n}$ is relevant to the prediction of the probabilities of the following $\sigma_n$. By passing on hidden states in Eq. \eqref{eq:softmax_layer} between sites, the RNN 
is capable of modeling strongly correlated distributions.
Hereafter, we shall call the dimension $d_h$ of the hidden state $\bm{h}_{n}$ the {\it number of memory units}.
We emphasize that the weights $W$ and $U$ and the biases $\bm{b}$ and $\bm{c}$ together comprise the variational parameters of our ansatz wave function of the next section.
These parameters are typically shared among the different values of $n$, giving rise to a highly compact parametrization of the probability distribution. 
Once the dimension $d_h$ is specified, the number of parameters in the ansatz is independent of the system size $N$.

By construction, the model allows for an efficient estimation of the normalized probability of a given configuration $\bm \sigma$. 
This construction is unlike energy-based models, which require intractable calculations of the partition function, or likelihood-free models such as Generative Adversarial Networks (GANs) that do not allow for an explicit estimation of probabilities~\cite{goodfellow2016nips,Goodfellow-et-al-2016}. The sequential process of computing the probability vectors $\bm{y}_n$ is schematically depicted in Fig.~\ref{fig:RNN}(a). 
Deep architectures can be obtained by stacking several RNN cells as shown in Fig.~\ref{fig:RNN}(b) for a general activation function $A$ (not necessarily Softmax).
As illustrated in Fig.~\ref{fig:RNN}(c), RNNs have the {\it autoregressive property}, meaning that the conditional 
probability $P(\sigma_n|\sigma_{<n})$ depends only on configurations 
$\sigma_1,\dots \sigma_{n-1}$. We also note that the computational cost
of sampling a configuration $\sigma_1,\dots\sigma_N$ is linear in the length $N$ of the configuration. 
Another important property of the normalized RNN probability distribution is that it can be used to produce successive samples
$\bm{\sigma}$ and $\bm{\sigma'}$ that are independent. 
Taking advantage of this property, the sampling procedure can be parallelized.

In practice, training vanilla RNNs can be challenging, since capturing long-distance correlations between the variables $\sigma_{n}$ tends to make the gradients either explode or vanish~\cite{Bengio94,Kolen2001,pascanu2013difficulty,chung2014}. Similar to MPS~\cite{fannes_finitely_1992}, long-distance correlations in RNNs are suppressed exponentially~\cite{shen2019} and extensions of the vanilla RNN have been proposed \cite{hochreiter1997long,cho-etal-2014-properties} in order to improve on this limitation. Two successful examples are the long short-term memory (LSTM) unit \cite{hochreiter1997long}, and the gated recurrent unit (GRU) \cite{cho-etal-2014-properties}. Unless stated otherwise, in this paper we use
the GRU \cite{cho-etal-2014-properties} as the elementary cell in our (one-dimensional) RNNs to study models in one and two spatial dimensions. The details of the implementation can be found in App.~\ref{sec:GRU}. 

Furthermore, we explore the use of two-dimensional (2D) vanilla RNNs~\cite{graves2012supervised},
where information about the spatial location of neighboring spins is exploited by the RNN ansatz.
The basic idea of 2D RNNs is to replace the single recurrent connection in a standard RNN, as shown in~\Eq{eq:rnn_action}, with two
recurrent connections that are passed to the neighboring sites. Thus, at each point in the
lattice the hidden layer of the network receives both spin configuration inputs and the hidden vectors
from the neighboring sites, in a way that respects the autoregressive property. 
We provide the details of the implementation in Sec.~\ref{sec:2D_TFIM} and App.~\ref{sec:2DRNN}.

\subsection{RNN wave functions}
\label{sec:RWF}
\begin{figure}[htb]
    \centering
    \includegraphics[width=0.95\linewidth]{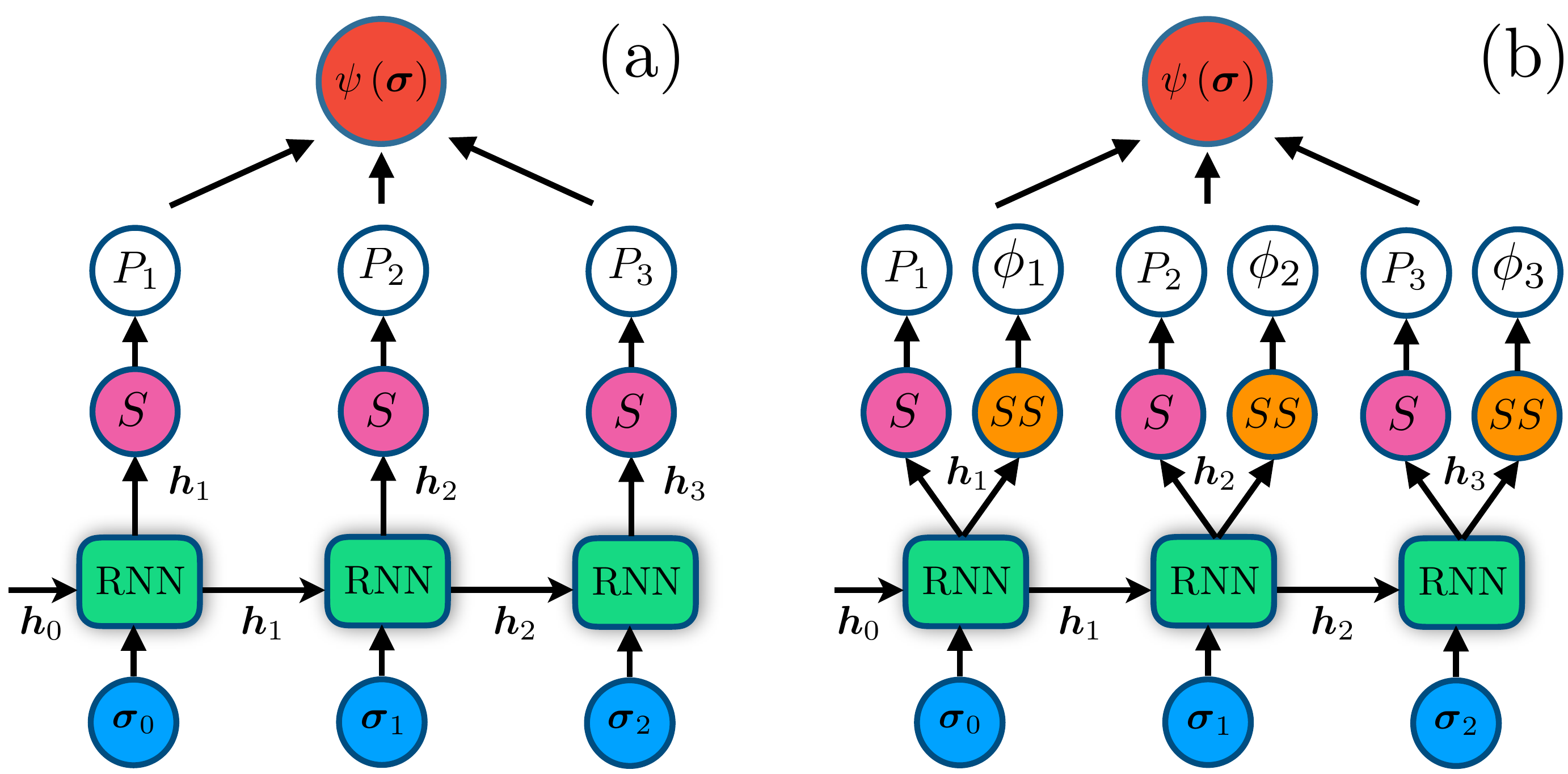}
    \caption{(a) pRNN wave function: A graphical representation of the computation of positive amplitudes using one RNN cell along with a Softmax layer (magenta circles) to compute the modulus $\lvert\psi(\bm{\sigma})\rvert^2 = P(\bm{\sigma})$.
    (b) cRNN wave function: A graphical representation of the computation of complex amplitudes using one RNN cell along with a Softmax layer (magenta circles) and a Softsign (SS) layer (orange circles). The first computes the modulus $\lvert\psi(\bm{\sigma})\rvert^2 = P(\bm{\sigma})$, the second to computes the phase $\phi(\bm{\sigma})$ of $\psi(\bm{\sigma})$.}
    \label{fig:RNNWF}
\end{figure}

The previous section focused exclusively on the efficient parametrization
of classical probability distributions $P(\bm{\sigma})$. In contrast,
quantum mechanical wave functions are in general a set of complex valued 
amplitudes $\psi(\bm{\sigma})$, rather than conventional probabilities.
Before discussing how to modify the RNN ansatz to represent complex wave functions,
we note that an important class of {\it stoquastic} many-body
Hamiltonians has ground states $\ket{\Psi}$ with real and positive amplitudes
in the standard product spin basis~\cite{Bravyi:2008:CSL:2011772.2011773}.
Thus, these ground states have representations in terms of probability distributions,
\begin{align}
    \ket{\Psi} = \sum_{\bm{\sigma}} \psi(\bm{\sigma})\ket{\bm{\sigma}} = \sum_{\bm{\sigma}}\sqrt{P(\bm{\sigma})}\ket{\bm{\sigma}}.
\end{align}
This property has been exploited extensively in wave function representations using generative models 
such as restricted Boltzmann machines~\cite{melko2019}.
For such wave functions, it is also natural to try to approximate $P(\bm{\sigma})$ with a
conventional RNN, as illustrated in Fig.~\ref{fig:RNNWF}(a). 
For later reference we call this architecture a {\it positive recurrent neural network wave function} (pRNN wave function). 

The generalization to the complex case starts by splitting the wave function into an amplitude and 
phase $\phi(\bm{\sigma})$ \cite{torlai2018} as
\begin{align}
    \ket{\Psi} = \sum_{\bm{\sigma}}\exp({\rm i}\phi(\bm{\sigma}))\sqrt{P(\bm{\sigma})}\ket{\bm{\sigma}}.
\end{align}
As illustrated in Fig.~\ref{fig:RNNWF}(b), we use one RNN cell and a Softmax layer to model the probability, 
together with a Softsign layer (as defined below) to model the phase.
In this parametrization, the first layer uses the Softmax activation function to get conditional probabilities $P_n$ as
\begin{align}
P_n = \bm{y}_n^{(1)} \cdot \bm{\sigma}_n,
\end{align}
where
\begin{equation}
\bm y^{(1)}_n = \text{S}\left(U^{(1)}\boldsymbol{h}_{n} + \bm{c}^{(1)}\right),
\end{equation}
in a similar fashion to Eq.~\eqref{eq:softmax_layer}. The Softsign layer is used to compute the phases as
\begin{align}
\phi_n = \bm{y}_n^{(2)} \cdot \bm{\sigma}_n,
\end{align}
where
\begin{equation}
\bm y^{(2)}_n = \pi\; \text{Softsign}\left(U^{(2)}\boldsymbol{h}_{n} + \bm{c}^{(2)}\right).
\end{equation}
The Softsign function is defined as
\begin{equation*}
    \text{Softsign}(x) = \frac{x}{1+\lvert x \rvert} \in (-1,1).
\end{equation*}
Finally, the probability $P(\bm{\sigma})$ is obtained from the $N$ individual
contributions $P_n$ as
\begin{align}
  P(\bm{\sigma}) \equiv \Pi_{n=1}^{N} P_n,
\end{align}
and, similarly, the phase $\phi(\bm \sigma)$ is computed as
\begin{align}
  \phi(\bm{\sigma}) \equiv \sum_{n=1}^{N} \phi_n.
\end{align}
Note that sampling from the square of the
amplitudes $P(\bm{\sigma})$ is unaffected by the Softsign layer and is
carried out, as described above, using only the Softmax layer as in Fig.~\ref{fig:RNN}(c).
For later reference, we call this architecture a {\it complex recurrent neural network wave function} (cRNN wave function), and hereafter, the term RNN wave function will refer to both pRNN wave functions and cRNN wave functions. 
Details about the dimensions of the variational parameters of RNN wave functions can be found in App.~\ref{sec:GRU}. 

%%%%%%%%%%%%%%%%%% RESULTS %%%%%%%%%%%%%%%%%%
%%%%%%%%%% TFIM %%%%%%%%%%
\section{Ground States with RNN wave functions}
\label{sec:results}
We focus our attention on the ground state properties of prototypical
Hamiltonians in condensed matter physics including the one- and two-dimensional (1D and 2D) transverse field Ising model (TFIM), as well as the 1D $J_1$-$J_2$ model, both with open boundary conditions. Their Hamiltonians are given by
\begin{equation}
    \hat{H}_{\text{TFIM}} = - \sum_{\langle i,j \rangle} \hat{\sigma}^{z}_i \hat{\sigma}^{z}_j - h \sum_{i} \hat{\sigma}^{x}_i,
    \label{eq:TFIM}
\end{equation}
where $\hat{\sigma}^{(x,y,z)}_i$ are Pauli matrices acting on site $i$, and 
\begin{equation}
    \hat{H}_{J_1-J_2} = J_1 \sum_{\langle i,j \rangle} \hat{{\bf S}}_i \cdot \hat{{\bf S}}_{j} + J_2 \sum_{\langle \langle i,j \rangle \rangle} \hat{{\bf S}}_i \cdot \hat{{\bf S}}_{j}.
    \label{eq:J1J2}
\end{equation}
where $\hat{{\bf S}}_i$ is a spin-1/2 operator.  Here,
$\langle i,j \rangle$ and $\langle \langle i,j \rangle \rangle$ denote
nearest- and next-nearest-neighbor pairs, respectively. Energies for the $J_1$-$J_2$
model are measured in units of $J_1=1$ in the results that follow.

To train our models we use the variational principle, where for a given problem Hamiltonian $\hat{H}$, the optimization strategy involves minimizing the expectation value $ E_\lambda = \braket{\Psi_\lambda|\hat{H}|\Psi_\lambda}\ge E_{0}$ with respect to the variational parameters $\lambda$. Here, $E_{0}$ is the exact ground state energy of $\hat{H}$. The variational parameters $\lambda$ are updated using variants of the gradient descent algorithm with the objective of minimizing $E_\lambda = \braket{\Psi_\lambda|\hat{H}|\Psi_\lambda}$. We provide a detailed description of the VMC scheme and the optimization strategy with which we optimize our RNN wave functions in App.~\ref{sec:VMC}.

Since the TFIM in \Eq{eq:TFIM} is stoquastic, the ground state is positive \cite{Bravyi:2008:CSL:2011772.2011773} and hence we use the pRNN wave function ansatz.
The $J_1$-$J_2$ model with positive couplings, on the other hand, has a ground state endowed with a sign 
structure in the computational $z$-basis, and thus we use a cRNN wave function ansatz. 

In the following sections, we use 1D RNN wave functions to approximate the ground state problem of the 1D TFIM and the 1D $J_1$-$J_2$ model, whereas we use both 1D and 2D pRNN wave functions in the case of the 2D TFIM.

\subsection{1D transverse field Ising model}
\label{sec:1DTFIMresults}

\begin{figure}[htp]
    \centering
    \includegraphics[width = \columnwidth]{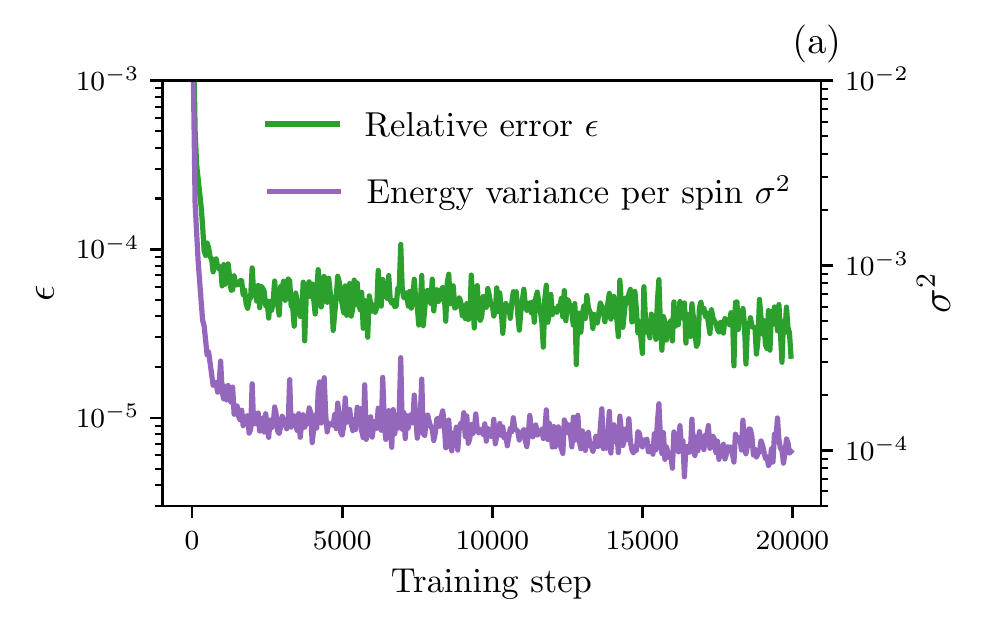}
    \includegraphics[width = \columnwidth]{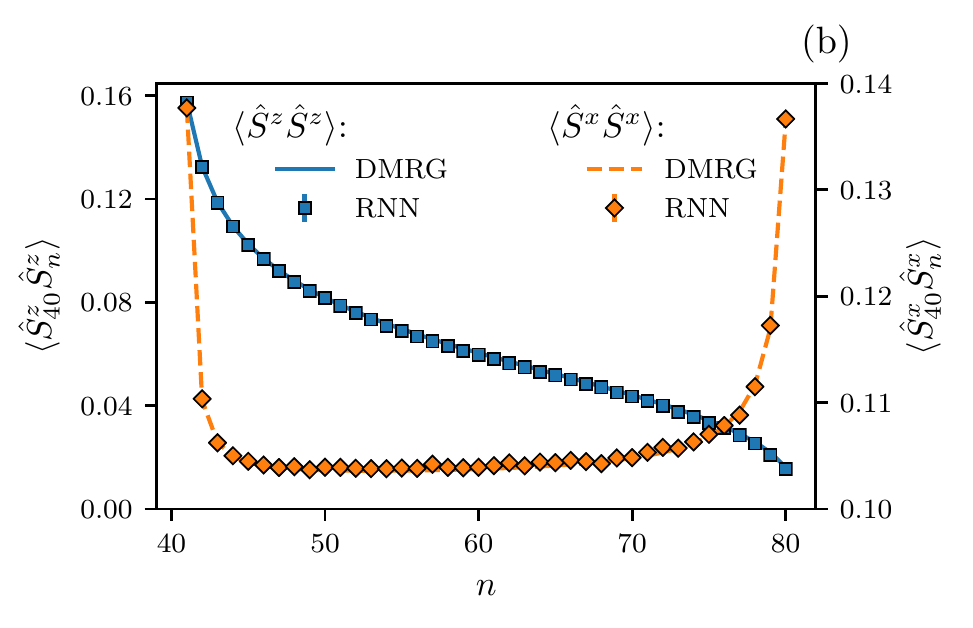}
    \includegraphics[width = \columnwidth]{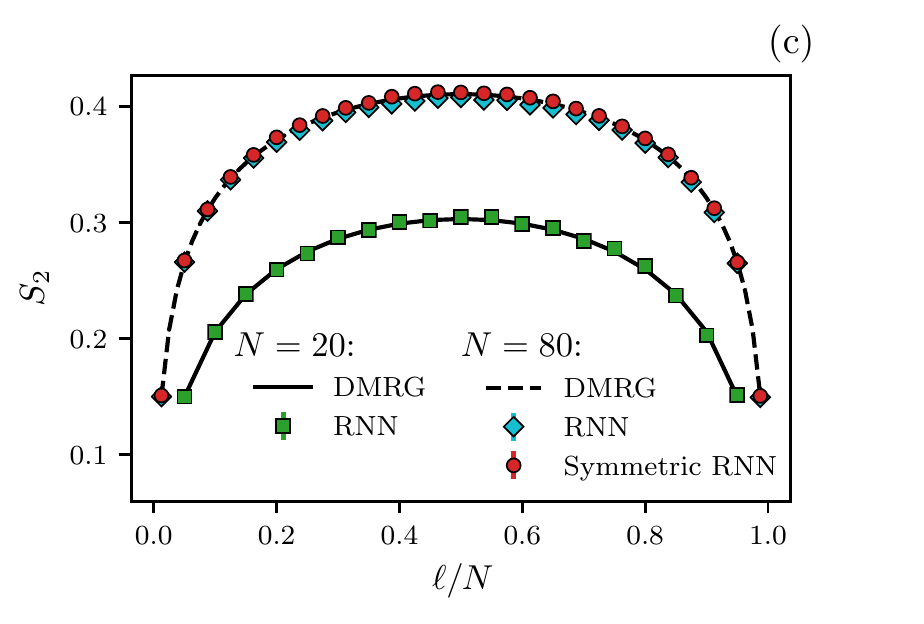}
    \caption{
    Results for the pRNN wave function compared with DMRG when targeting the ground state of a 1D TFIM at the critical point. 
    Our pRNN wave function has one layer with 50 units.
    (a) The relative error $\epsilon$ and the energy variance per spin $\sigma^2$ against the number of training steps (i.e. gradient descent steps) for $N=1000$ spins. We use only 200 samples per gradient step, which are enough to achieve convergence. 
    (b) The two-point correlation function $\langle \hat{S}_{40}\hat{S}_n \rangle$ along the $x$-axis and $z$-axis of the optimized pRNN wave function for sites $n>40$ using $10^6$ samples. DMRG results are also shown for comparison. (c) The R\'enyi entropy $S_2$ against the relative size of subregion $A$ for system sizes $N=20$ and $80$. 
    In both~(b) and~(c), the error bars are smaller than the data points.
    }
    \label{fig:1d_TFIM_results}
\end{figure}

To demonstrate the power of our proposed method, we
use it to target the ground state of a TFIM in one dimension with $N=1000$
spins at the critical point $h = 1$ using a pRNN wave function that has a single-layer RNN with $50$ memory units. In \Fig{fig:1d_TFIM_results}(a), we show the evolution of the relative error
\begin{equation}
    \epsilon \equiv \frac{|E_{\rm RNN} - E_{\rm DMRG}|}{|E_{\rm DMRG}|},
\end{equation}
and the energy variance per spin
\begin{equation}
    \sigma^2 \equiv \frac{\left\langle \hat{H}^2 \right\rangle - \left\langle \hat{H} \right\rangle^2}{N},
    \label{eq:variance}
\end{equation}
as a function of the training step. $E_{\text{DMRG}}$ is the ground state energy as obtained from a density matrix renormalization group (DMRG) calculation~\cite{white_density_1992,roberts2019tensornetwork}, and can be considered exact in one dimension. We obtain very accurate results with a modest number of parameters ($\sim 8000$, see App.~\ref{sec:GRU}). 
For comparison, the number of parameters of a restricted Boltzmann machine (RBM) \cite{carleo2017} with one layer scales as $M N$ with $M$ the number of hidden units and $N$ the number of physical spins.  This scaling implies that the pRNN wave function here has the same number of variational parameters as an RBM with only eight hidden units. 

While energies and variances give a quantitative indication of the quality of a variational wave function, correlation functions provide a more comprehensive characterization. Indeed, correlation functions are at the heart of condensed matter theory since many experimental probes in condensed matter physics directly relate to measurements of correlation functions. Examples include inelastic scattering, which probes density-density correlation functions, and the Green's function, out of which important thermodynamic properties of a quantum system can be computed~\cite{Abrikosov:107441}. In \Fig{fig:1d_TFIM_results}(b)  we compare the RNN results for the two-point correlation functions $\braket{\hat S^x_n\hat S^x_m}$ and $\braket{\hat S^z_n\hat S^z_m}$ with DMRG. 
Here, we see consistency between the RNN and the DMRG results.

Extracting entanglement entropy from many-body quantum systems is a central theme in condensed matter physics, with entanglement entropy providing an additional window into the structure of complex quantum states of matter beyond what is seen from correlation functions. Of particular interest is the family of {\it R\'enyi entropies}
of order $\alpha$ of a reduced density matrix $\rho$,
\begin{equation}
  S_\alpha (\rho) = \frac{1}{1-\alpha} \log \left ( \text{Tr} \rho^\alpha \right ).
\end{equation}
$S_\alpha (\rho)$ encodes important non-local properties of quantum many-body systems such as topological entanglement, and contains information about universal properties of
quantum phases such as the central charge $c$~\cite{Alioscia2009, Ryu2006}. Due to their
non-local character, extracting R\'enyi entropies from many-body quantum systems is notoriously
difficult. Here, we use the so-called {\it replica trick} \cite{hastings2010measuring} to calculate
the $\alpha=2$ R\'enyi entropy $S_2(\rho)$ for RNN wave functions. The details of the implementation can be found in App.~\ref{sec:entropy}. In \Fig{fig:1d_TFIM_results}(c), we show results for the R\'enyi entropy $S_2(\rho_\ell)$
for two different system sizes $N=20,80$ of 1D TFIM. $\rho_\ell$ here is the reduced density matrix
on the first $\ell$ sites of the spin chain, obtained by tracing out all sites $n \in [\ell+1, L]$ such that
\begin{align}
  \rho_\ell = \tr_{n\in [\ell+1,L]}\left(\ket{\Psi}\bra{\Psi}\right).
\end{align}
Indeed for both system sizes, \Fig{fig:1d_TFIM_results}(c) shows excellent agreement between the pRNN wave function estimation and the DMRG result. To improve the overall quality of the quantum state,  we have enforced the parity symmetry on our pRNN wave function (see App.~\ref{sec:discrete_symmetries}), denoted by ``Symmetric RNN'' in \Fig{fig:1d_TFIM_results}(c). We observe that the symmetric pRNN wave function leads to a more accurate estimate of $S_2(\rho_\ell)$ for $N=80$ sites.

%%%%%%%%%% Frustrated J1-J2 %%%%%%%%%%%%%%%%%%%%%%%%%%%%%%%%%%%%%%%%%%%%%%%%%%%%%%%%%%%%%%%%%%%%%%%%%%%%%%%
\subsection{1D $J_1-J_2$ model}
Moving beyond stoquastic Hamiltonians,  we now investigate
the performance of RNN wave functions for a Hamiltonian
the ground state of which has a sign structure in the computational basis, specifically the $J_1$-$J_2$ model in one dimension.

We use a variationally optimized deep cRNN wave function with three GRU layers, each with 100 memory units, to approximate the ground state of the $J_1$-$J_2$ model.
The phase diagram of this model has been studied with DMRG~\cite{white1996}, where it was found that the model exhibits a quantum phase transition at $J_2^c = 0.241167 \pm 0.000005$ \cite{Eggert1996, becca2009variational}
from a critical Luttinger liquid phase for $J_2 \leq J_2^c$
to a spontaneously dimerized gapped valence bond state phase for $J_2 \geq J_2^c$.

We impose $U(1)$ spin symmetry in the cRNN wave function
(see App.~\ref{sec:zeromag}), and target the ground state at four different points $J_2 = 0.0,0.2,0.5,0.8$.
Note that at $J_2=0$, the Hamiltonian in \Eq{eq:J1J2} can be made stoquastic by a local unitary transformation that rotates every other spin by $\pi$ around the $z$-axis. The ground state can in this case be decomposed as \cite{Marshall1955}
\begin{equation}\label{eq:marshall}
    \psi({\bm \sigma}) = (-1)^{M_A({\bm \sigma})} \tilde \psi({\bm \sigma}),
\end{equation}
where $M_A(\bm \sigma)$ is given by $M_A(\bm{\sigma}) = \sum_{ i\in A} \sigma_i$ with $\sigma_i \in \{0,1\}$ \cite{Marshall1955} and $\tilde \psi({\bm \sigma})$ is the {\it positive} amplitude of the wave function. The set $A$ comprises the sites belonging to the sublattice of all even (or all odd) sites in the lattice. 
The prefactor $(-1)^{M_A({\bm \sigma})}$ is known as the Marshall sign of the wave function~\cite{Marshall1955}.
For $J_2\neq 0$, this decomposition is no longer exact, and $\tilde \psi(\bm \sigma)$ acquires 
a non-trivial sign structure. For finite $J_2$ the decomposition in \Eq{eq:marshall} can still be
applied with the hope that the sign structure of $\psi(\bm \sigma)$ remains close to the Marshall sign~\cite{choo2019study}.

\begin{figure}[tb]
    \centering
    \includegraphics{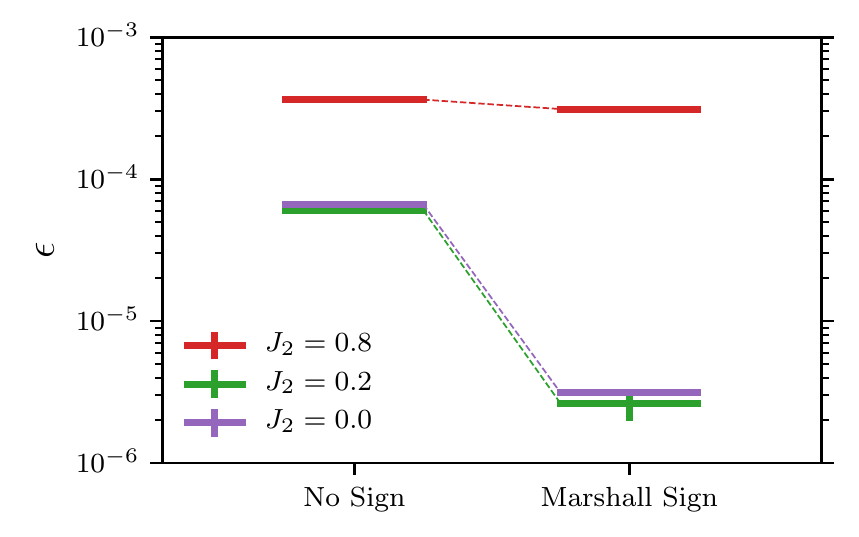}
    \caption{The relative error (compared to DMRG) of the cRNN wave function trained on the 1D $J_1-J_2$ model with $N=100$ spins for different values $J_2$, 
    both without a prior sign (represented by ``No Sign'') and with a prior Marshall sign as in \Eq{eq:marshall} (represented by ``Marshall Sign'').
    We observe that applying a Marshall sign improves the accuracy.
    }
    \label{fig:J1J2_results}
\end{figure}

In \Fig{fig:J1J2_results}, we compare ground state energies of the cRNN wave function trained on the 1D $J_1$-$J_2$ model with $N=100$ spins with and without applying a Marshall sign. 
For small values of $J_2$, we find a considerable improvement of the energies when applying the Marshall sign on top of the cRNN wave function. This observation highlights the importance of considering a prior ``sign ansatz'' to achieve better results. 
In the absence of a prior sign, the cRNN wave function can still achieve accurate estimations of the ground state energies, showing that cRNN wave functions can recover some of the unknown sign structure of the ground state. For $J_2=0.8$, however, the improvement is less pronounced, which is expected due to the emergence of a second sign structure in the limit $J_2 \to \infty$ (when the system decouples into two independent unfrustrated Heisenberg chains)~\cite{torlai2019a, Thibaut_2019}, that is widely different from the Marshall sign in Eq.~\eqref{eq:marshall}.
We omit from \Fig{fig:J1J2_results} our results at the point $J_2=0.5$. In this case, the 1D $J_1$-$J_2$ model reduces to the Majumdar-Ghosh model, where the ground state is a product-state of spin singlets, and we find agreement with the exact ground state energy within error bars when we apply an initial Marshall sign structure.
We provide a summary of the cRNN wave function's obtained values in App.~\ref{sec:tables_of_results}. 

%%%%%%%%%%%%%%%%%%%%%%%%%%%%%%%%%%%%%%%%%%%%%%%%%%%%%%%%%%%%%%%%%%%%%%%%%%%%%%%%%%%%%%%
%%%%%%%%%%%%%%%%%%%%%%%%%%%%%   2D Ising     %%%%%%%%%%%%%%%%%%%%%%%%%%%%%%%%%%%%%%%%%%
%%%%%%%%%%%%%%%%%%%%%%%%%%%%%%%%%%%%%%%%%%%%%%%%%%%%%%%%%%%%%%%%%%%%%%%%%%%%%%%%%%%%%%%

\subsection{2D transverse field Ising model}
\label{sec:2D_TFIM}
\begin{figure*}[!htp]
    \centering
    \includegraphics[width = 0.9\linewidth]{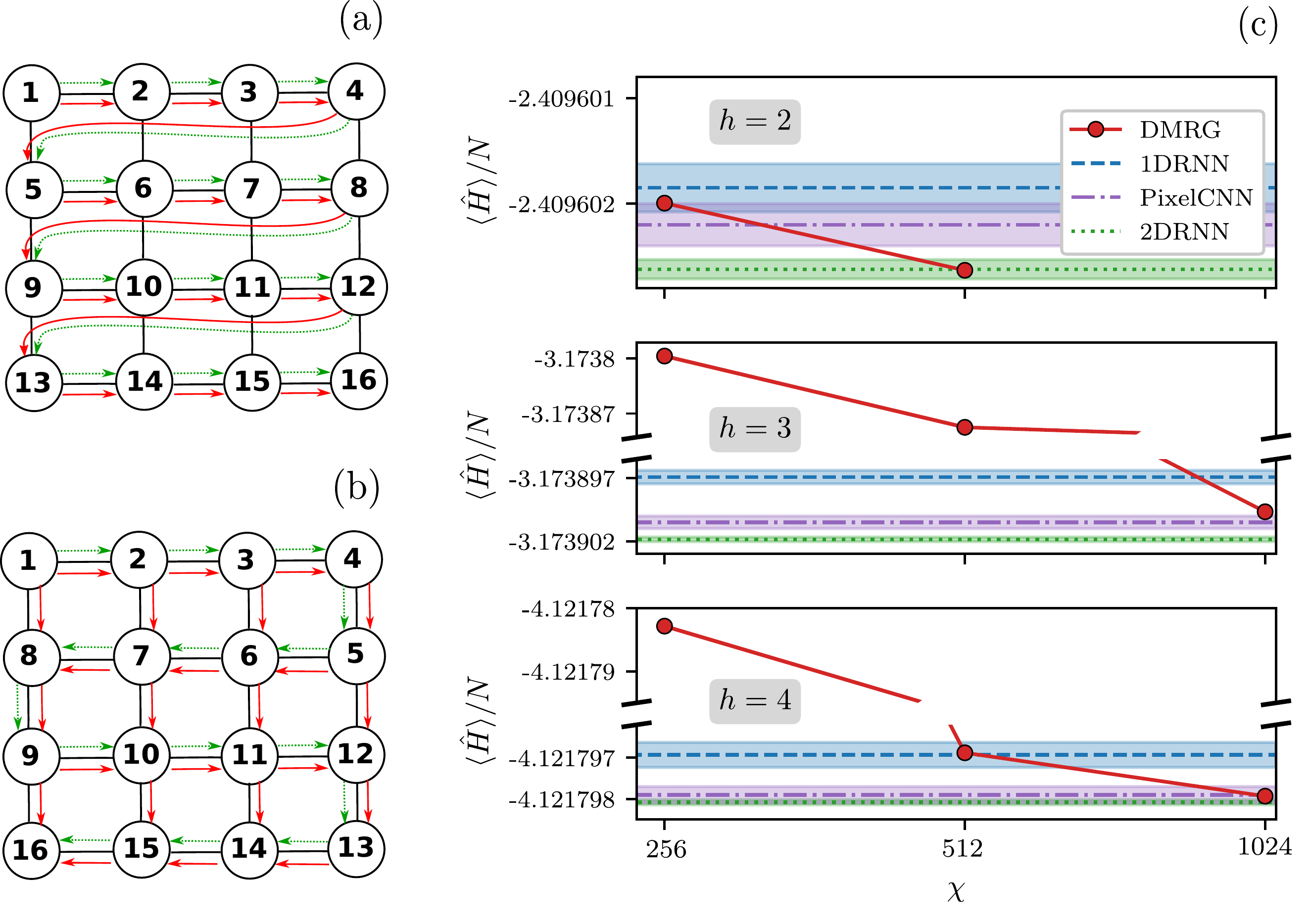}
    \caption{ (a): Autoregressive sampling path of 2D spin configurations using 1D RNN wave functions.
      The 2D configurations are generated through raster scanning, such that in order to generate spin $\sigma_i$ one has to condition on the spins that are previously generated. (b): Autoregressive sampling path of 2D spin configurations using 2D RNN wave functions through a zigzag path, where each site receives two hidden states and two spins from the horizontal and the vertical neighbors that were previously generated. For both panels (a) and (b), the digits and the green dashed arrows indicate the sampling path, while the red arrows indicate how the hidden states are passed from one site to another. (c): A comparison of the variational energy per spin between a 2D pRNN wave function (labeled as 2DRNN), 1D pRNN wave function (labeled as 1DRNN),
      PixelCNN wave function~\cite{sharir2019deep}, and DMRG with bond dimension $\chi$ for the 2D TFIM on a system with $L_x \times L_y=12 \times 12$ spins. The shaded regions represent the error bars of each method.
      Note the broken $y$-axis on the plots for $h=3$ and $4$, denoting a change in scale between the upper and lower portions of the plots. These results show that 2D pRNN wave functions can achieve a performance comparable to PixelCNN wave functions and DMRG with a large bond dimension, while requiring only a fraction of their variational parameters. 
      }
    \label{fig:2DTFIM_study}
\end{figure*}
Understanding strongly correlated quantum many-body systems in $D>1$ spatial dimensions
is one of the central problems in condensed matter physics. During the last decade,
numerical approaches such as tensor networks~\cite{verstraete_renormalization_2004,yan_spin-liquid_2011, evenbly_tensor_2015},
quantum Monte Carlo~\cite{sandvik2010,becca2017}, and neural networks~\cite{carleo2017} have moved
to the forefront of research in this area. Despite tremendous progress, however, solving correlated quantum many-body systems even in two dimensions remains a challenging problem. We now turn our attention to the application of our
RNN wave function approach to the 2D quantum Ising model shown in \Eq{eq:TFIM} on a 
square lattice, a paradigmatic example of a strongly correlated quantum many-body system.
This model has a quantum phase transition at a critical magnetic field $h^c\approx 3.044$ that separates a magnetically ordered phase from a random paramagnet~\cite{PhysRevE.66.066110}.

The simplest strategy for extending our approach to 2D geometries is to simply treat them as
folded 1D chains, similar to the ``snaking'' approach used in 2D DMRG calculations (see \Fig{fig:2DTFIM_study}(a)). While this approach works quite well, it has the fundamental drawback that neighboring
sites on the lattice can become separated in the 1D geometry. As a consequence,
local correlations in the 2D lattice are mapped into non-local correlations in the 1D geometry,
which can increase the complexity of the problem considerably.
For example, 2D DMRG calculations are typically restricted to 2D lattices with small width $L_y$. This problem has led to the development of more powerful tensor network algorithms for 2D quantum systems such as projected entangled pair states (PEPS)~\cite{verstraete_renormalization_2004}.

An advantage of RNN wave functions is their flexibility in how hidden vectors
are passed between units. To obtain an RNN wave function more suited to a 2D geometry,
we modify the simple 1D approach outlined above by allowing hidden vectors to also be passed vertically,
instead of only horizontally, as described in App.~\ref{sec:2DRNN}. This modification is illustrated by the red arrows in \Fig{fig:2DTFIM_study}(b). We refer to
this geometry in the following discussions as a 2D RNN.
We optimize the 2D pRNN wave function with a single-layer 2D vanilla RNN cell that has 100 memory units
(i.e. with $\sim 21000$ variational parameters) to approximate the ground state of
the 2D quantum Ising model at $h = {2,3,4}$. 
The training complexity
of the 2D pRNN wave function is only quadratic in the number of
memory units $d_h$ (see App.~\ref{sec:2DRNN}), which is very inexpensive compared to, e.g.,
the expensive variational optimization of PEPS, which scales as $\chi^2\tilde{D}^6$ 
(where $\tilde{D}$ is the PEPS bond dimension and $\chi$ is the bond dimension of the intermediate MPS)~\cite{vanderstraeten_gradient_2016}.

For comparison, we also optimize a deep 1D pRNN wave function architecture with three layers of
stacked GRU cells, each with 100 memory units (i.e., with $\sim$152000 variational parameters)
for the same values of the magnetic field $h$.
In \Fig{fig:2DTFIM_study}(c) we compare the obtained ground state energies with results from
2D DMRG calculations (run on the same 1D geometry as for the 1D pRNN wave function) and the PixelCNN
architecture~\cite{PixelCNN} (with $\sim$800000 variational parameters and results are taken from Ref.~[\onlinecite{sharir2019deep}]).
For the magnetic fields shown above and for large bond dimensions,
we obtain excellent agreement between all four methods.
This agreement is particularly remarkable given that the 2D pRNN wave function uses only about 0.03\%
of the variational parameters of the DMRG calculation with bond dimension $\chi = 512$, 
about 2.6\% of the variational parameters of the PixelCNN wave function used in Ref.~[\onlinecite{sharir2019deep}],
and about 14\% of the parameters used in the 1D pRNN architecture.
A summary of our results in tabular form can be found in App.~\ref{sec:tables_of_results}.

%%%%%%%%%%%%%%%%%%%%%%%%%%%%%%%%%%%%%%%%%%%%%%%%%%%%%%%%%%%%%%%%%%%%%%%%%%%%%%%%%%%%%%%
%%%%%%%%%%%%%%%%%%%%%%%%%%%%%%%%%%%%%%%%%%%%%%%%%%%%%%%%%%%%%%%%%%%%%%%%%%%%%%%%%%%%%%%
%%%%%%%%%%%%%%%%%%%%%%%%%%%%%%%%%%%%%%%%%%%%%%%%%%%%%%%%%%%%%%%%%%%%%%%%%%%%%%%%%%%%%%%
\subsection{Scaling of resources}

\begin{figure}[!htp]
    \centering
    \includegraphics[width = \columnwidth]{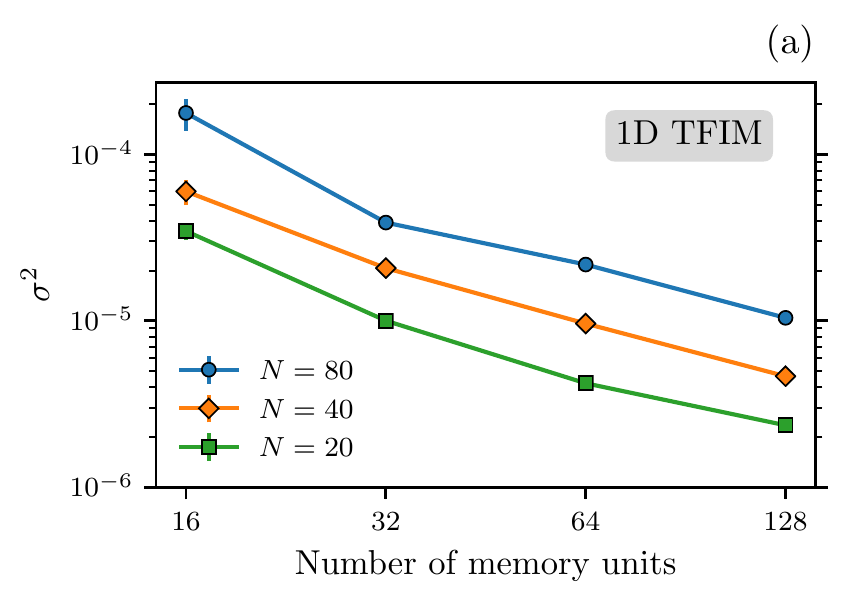}
    \includegraphics[width = \columnwidth]{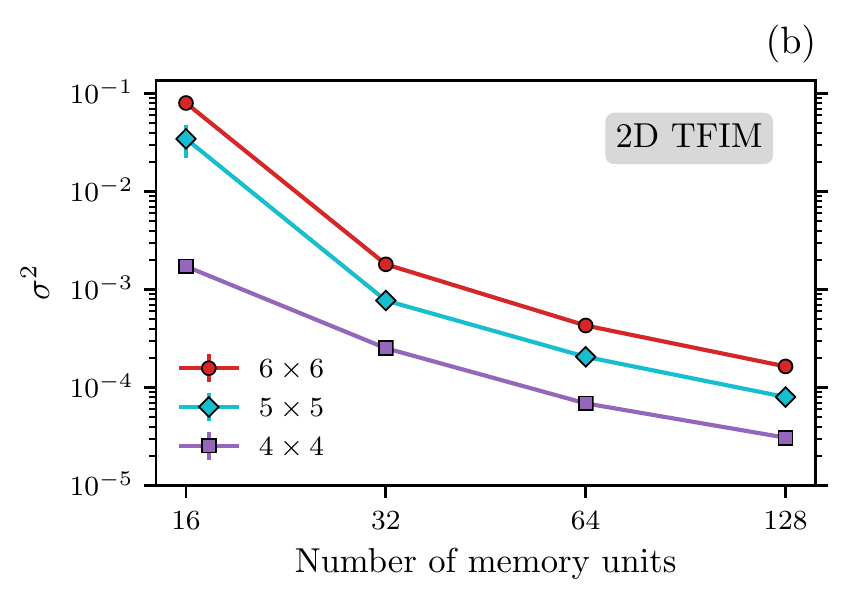}
    \caption{The energy variance per spin against the number of memory units of a 1D pRNN wave function trained at the critical point of (a) the 1D TFIM and (b) the 2D TFIM. 
    Both scalings show that we can systematically reduce the bias in the estimation of the ground state energy.}
    \label{fig:Scaling_study}
\end{figure}

The optimization results of our RNN wave function approach depend on several hyperparameters, including the number of memory units, the number of recurrent layers in deep architectures, and the number of samples used to obtain the gradient during an optimization step (see App.~\ref{sec:VMC}). Here, we investigate how the optimized energy and the energy variance per spin $\sigma^2$ (see \Eq{eq:variance}) depend on these parameters.
This energy variance per spin is an indicator of the quality of the optimized wave function, with exact eigenstates corresponding to $\sigma^2=0$. When targeting eigenstates, deviations from this value can be used to assess the quality of a variational wave function~\cite{Claudius90,Assaraf2003,becca2017}, as previously done in the case of matrix product state based techniques~\cite{Hubig_2018,bauls2019entanglement}. For variational approaches such as DMRG, one typically expects a non-zero value of $\sigma^2$ that decreases when one increases the number of parameters (i.e.,~the expressivity) of the variational wave function. Since the number of variational parameters is directly related to the number of memory units of the pRNN wave function (see App.~\ref{sec:GRU}), we study here the scaling of $\sigma^2$ with the number of memory units. 

In \Fig{fig:Scaling_study}, we present the dependence of $\sigma^2$ on the number of memory units for the 1D and 2D critical TFIMs.
\Fig{fig:Scaling_study}(a) shows results for $\sigma^2$ for a 1D critical TFIM on three system sizes $N=20, 40$ and $80$, and \Fig{fig:Scaling_study}(b) shows results for the 2D TFIM
on $4\times 4, 5\times 5$ and $6\times 6$ square lattices. In all cases, we used a single-layer 1D pRNN wave function and $500$ samples during optimization to compute estimates of the gradients. For each $N$ we observe a systematic decrease of $\sigma^2$ (i.e., an increase in quality of the wave function) as we increase the number of memory units.

In App.~\ref{sec:scaling_app}, we study the dependence of $\sigma^2$ on both the number of samples and the number of layers in the pRNN wave function for a critical 1D TFIM.
We observe only a weak dependence on both parameters.
The weak dependence on the number of samples suggests that optimizing the RNN wave functions with noisy gradients does not significantly impact the results of the optimization procedure, and yields accurate estimations of the ground state and its energy.
From the weak dependence on the number of layers we conclude that shallow RNNs with a sufficient number of memory units have enough expressivity and that deep architectures do not seem to be beneficial from an accuracy point of view.
However, deeper networks could have potential ramifications regarding memory usage and training speed when it comes to training a large number of variational parameters, as shallow RNNs with a large number of memory units are equivalent in terms of number of parameters to deep RNNs with a smaller number of memory units. 
We also note that adding residual connections between layers \cite{hermans2013training} and dilated connections between RNN cells \cite{chang2017dilated} to deep RNNs, which we leave for future investigations, might change our previous conclusions and make deep RNNs more beneficial compared to shallow RNNs.

%%%%%%%%%%%%%%%%%%%%%%%%%%%%%%%%%%%%%%%%%%%%%%%%%%%%%%%%%%%%%%%%%%%%%%%%%%%%%%%%%%%%%%%
%%%%%%%%%%%%%%%%%%%%%%%%%%%%%%%%%%%%%%%%%%%%%%%%%%%%%%%%%%%%%%%%%%%%%%%%%%%%%%%%%%%%%%%
%%%%%%%%%%%%%%%%%%%%%%%%%%%%%%%%%%%%%%%%%%%%%%%%%%%%%%%%%%%%%%%%%%%%%%%%%%%%%%%%%%%%%%%

%%%%%%%%%%%%%%%%%% RESULTS %%%%%%%%%%%%%%%%%%
\section{Conclusions and outlook}
We have introduced recurrent neural network wave functions, a novel variational ansatz for quantum many-body systems, which we use to approximate ground state energies, correlation functions, and entanglement of many-body Hamiltonians of interest to condensed matter physics.
We find that RNN wave functions are competitive with state-of-the-art methods such as DMRG and PixelCNN wave functions~\cite{sharir2019deep}, performing particularly well on the task of finding the ground state of the transverse-field Ising model in two dimensions. By increasing the number of memory of units in the RNN, the error in our results can be systematically reduced. 
We have shown furthermore that we can accurately model ground states endowed with a sign structure using a complex recurrent neural network (cRNN) wave function ansatz. Here, accuracy can be improved by introducing an ansatz sign structure and by enforcing symmetries such as $U(1)$ symmetry. 
The autoregressive nature of RNN wave functions makes it possible 
to directly generate independent samples, in contrast to methods based on Markov chain sampling, which are often plagued by long autocorrelation times that affect the optimization and
the accurate estimation of correlation functions in a variational ansatz. 
Thanks to weight sharing among lattice sites, RNN wave functions provide very compact yet expressive representations of quantum states, while retaining the ability to easily train with millions of variational parameters, as opposed to, e.g., restricted Boltzmann machines~\cite{carleo2017}.
We expect that future work incorporating additional numerical techniques such as attention \cite{bahdanau2014neural,Transformer} and higher order optimization \cite{becca2017,martens2018kronecker} will make RNN wave functions a highly competitive tool for simulating quantum many-body systems, with applications to material science, quantum chemistry~\cite{choo2019fermionic}, quantum computation~\cite{carrasquilla2019probabilistic}, and beyond.

{\it Note added}. A complementary paper on recurrent neural network wave functions~\cite{roth2020iterative} appeared after the publication of this manuscript.

%%%%%%%%%%%%%%%%%% Open-Source Code %%%%%%%%%%%%%%%%%%

\section*{Open-Source Code}
Our code is made publicly available at ``\url{http://github.com/mhibatallah/RNNwavefunctions}''. The hyperparameters we use are given in App.~\ref{sec:hyperparams}.

%%%%%%%%%%%%%%%%%% ACKNOWLEDGMENTS %%%%%%%%%%%%%%%%%%

\section*{Acknowledgments}

We acknowledge Di Luo for his generous comments which were extremely helpful. We also thank Estelle Inack, Dan Sehayek, Amine Natik, Matthew Beach, Bohdan Kulchytskyy, Florian Hopfmueller, Roeland Wiersema, Giuseppe Carleo and Noam Wies for useful discussions and insights. M.H. acknowledges support of the Ecole Normale Superieure de Lyon. M.G. acknowledges support by the Simons Foundation 
(Many Electron Collaboration). 
R.G.M. acknowledges support from the Natural Sciences and Engineering Research Council of Canada (NSERC), a Canada Research Chair, the Shared Hierarchical Academic Research Computing Network (SHARCNET) and Compute Canada.
J.C. acknowledges support from NSERC, SHARCNET, Compute Canada, and the 
Canadian institute for advanced research (CIFAR) AI chair program. Computer simulations were also made possible 
thanks to the Vector Institute computing cluster and Google Colaboratory.
This research was supported in part by the National Science Foundation 
under Grant No. NSF PHY-1748958. It was also supported in part by the Perimeter Institute for Theoretical Physics. Research at Perimeter Institute is supported in part by the Government of Canada through Innovation, Science and Economic Development Canada (ISED) and by the Province of Ontario through the Ministry of Economic Development, Job Creation and Trade.

%%%%%%%%%%%%%%%%%% APPENDICES %%%%%%%%%%%%%%%%%%
\appendix 

\section{Gated Recurrent Neural Networks}
\label{sec:GRU}

We use the GRU model introduced in Ref.~[\onlinecite{cho-etal-2014-properties}], which
processes the spin configurations $\bm{\sigma}$ as
\begin{align}
\label{eq:GRU}
\boldsymbol{u}_n &=\textrm{sig}\left( W_u\left[ \boldsymbol{h}_{n-1};\boldsymbol{\sigma}_{n-1}\right] + \bm{b}_u\right),  \\
\nonumber
\boldsymbol{r}_n &=\textrm{sig}\left(  W_r\left[ \boldsymbol{h}_{n-1};\boldsymbol{\sigma}_{n-1}\right] + \bm{b}_r\right),  \\ \nonumber
\boldsymbol{\tilde{h}}_n &= \tanh\left( W_c  \left[\boldsymbol{r}_n \odot \boldsymbol{h}_{n-1};\boldsymbol{\sigma}_{n-1} \right] + \bm{b}_c\right),\\ \nonumber
\boldsymbol{h}_{n} & = (1-\boldsymbol{u}_n)\odot\boldsymbol{h}_{n-1} + \boldsymbol{u}_n \odot \boldsymbol{\tilde{h}}_n,
\end{align}
where sig and tanh represent the sigmoid and hyperbolic tangent activation functions, respectively.
Thus, the hidden vector $\boldsymbol{h}_{n}$ is updated through an interpolation between the previous
hidden state $\boldsymbol{h}_{n-1}$ and a candidate hidden state $\boldsymbol{\tilde{h}}_n$. 
The update gate $\boldsymbol{u}_n$ decides to what extent
the contents of the hidden state are modified, and depends on how relevant the
input $\bm{\sigma}_{n-1}$ is to the prediction (Softmax layer).
The symbol $\odot$ denotes the pointwise (Hadamard) product.
The reset gate modeled by the vector $\boldsymbol{r}_n$ is such
that if the $i$-th component $\boldsymbol{r}_n$ is close to zero,
it cancels out the $i$-th component of the hidden vector state $\bm{h}_{n-1}$, effectively making the GRU ``forget'' part of the sequence that has
already been encoded in the state vector $\boldsymbol{h}_{n-1}$. 

\begin{figure}[tb]
\centering
\includegraphics[width=0.8\linewidth]{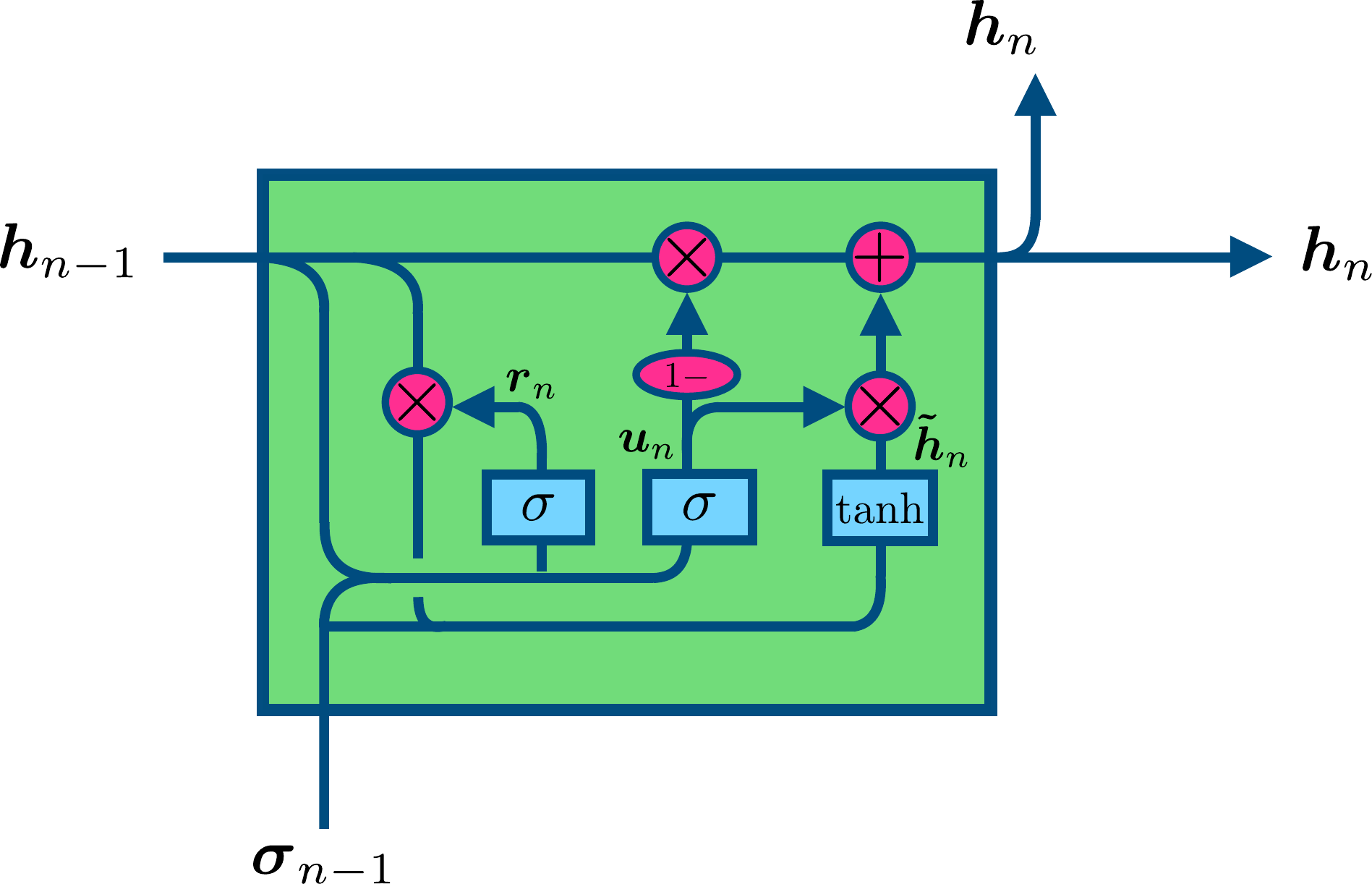}
\caption{
Graphical representation of the gated recurrent unit cell described in Eq.~\eqref{eq:GRU}. The magenta circles/ellipses represent point wise operations such as vector addition or multiplication. The blue rectangles represent neural network layers labeled by the non-linearity we use. Merging lines denote vector concatenation and forking lines denote a copy operation. 
The sigmoid activation function is represented by $\sigma$.}
\label{fig:GRU_RNN}
\end{figure}

The weights matrices $W_{u,r,c}$ and the bias vectors $\bm{b}_{u,r,c}$ parametrize the
GRU and are optimized using energy minimization as described in App.~\ref{sec:VMC}.
The GRU transformations in Eq.~\eqref{eq:GRU} are depicted graphically in Fig.~\ref{fig:GRU_RNN}.

To take advantage of the GPU speed up, we use instead the cuDNN variant of GRUs implemented in Tensorflow~\cite{tensorflow2015-whitepaper}, with
\begin{align}
\label{eq:GRU1}
\boldsymbol{u}_n &=\textrm{sig}\left( W_u\left[ \boldsymbol{h}_{n-1};\boldsymbol{\sigma}_{n-1}\right] + \bm{b}_u\right),  \\
\nonumber
\boldsymbol{r}_n &=\textrm{sig}\left(  W_r\left[ \boldsymbol{h}_{n-1};\boldsymbol{\sigma}_{n-1}\right] + \bm{b}_r\right),  \\ \nonumber
\boldsymbol{h'}_{n} &= W_c^{(1)} \boldsymbol{h}_{n-1} + \bm{b}_c^{(1)}, \\ \nonumber
\boldsymbol{\tilde{h}}_n &= \tanh\left( W_c^{(2)}  \boldsymbol{\sigma}_{n-1} + \boldsymbol{r}_n \odot \boldsymbol{h'}_{n} + \bm{b}_c^{(2)}\right), \\ \nonumber
\boldsymbol{h}_{n} & = (1-\boldsymbol{u}_n)\odot\boldsymbol{h}_{n-1} + \boldsymbol{u}_n \odot \boldsymbol{\tilde{h}}_n,
\end{align}
which differs slightly from the above implementation of traditional GRU cells~\cite{appleyard2016optimizing}.

Provided that the dimensions of the hidden state $\bm{h}_{n-1}$ and input $\bm{\sigma}_{n-1}$ are $d_h$ and $d_v$ (respectively), then the dimensions of the variational parameters of a GRU as in Eq.~\eqref{eq:GRU1} are
\begin{itemize}
    \item ${\rm dim} (W_{u,r}) =  d_h \times (d_h +d_v)$,
    \item ${\rm dim} (\bm{b}_{u,r}) = d_h$,
    \item ${\rm dim} (W_{c}^{(1)}) = d_h \times d_h$,
    \item ${\rm dim} (W_{c}^{(2)}) = d_h \times d_v $,
    \item ${\rm dim} (\bm{b}_c^{(1,2)}) = d_h$.
\end{itemize}
The new hidden state $\bm{h}_n$ is fed into a Softmax layer to infer conditional probabilities, such that
\begin{equation*}
\bm{y^}{(1)}_n = {\rm Softmax}(U^{(1)} \bm{h}_n + \bm{c}^{(1)}),
\end{equation*}
and also into a Softsign layer to infer the phases as
\begin{equation*}
\bm{y}^{(2)}_n = \pi \ {\rm Softsign}(U^{(2)} \bm{h}_n + \bm{c}^{(2)}).
\end{equation*}
We require the outputs $\bm{y^}{(1,2)}_n$ to have dimension $d_v$, so that each element of $\bm{y^}{(1)}_n$ represents the conditional probability of sampling a value for the next spin $\sigma_n \in \{0,1, \ldots, d_v-1\}$, and that each element of $\bm{y^}{(2)}_n$ corresponds to the phase of the chosen spin $\sigma_n$. Thus, the dimension of the parameters introduced in the Softmax/Softsign layer are
\begin{itemize}
    \item ${\rm dim} (U^{(1,2)}) = d_v \times d_h$,
    \item ${\rm dim} (\bm{c}^{(1,2)}) = d_v$.
\end{itemize}
The same reasoning can be also applied to determine the dimensions of the variational parameters of 2D vanilla RNNs presented in App.~\ref{sec:2DRNN}.

% ////////////////////////////////////////

\section{Two-dimensional Recurrent Neural Network wave functions}
\label{sec:2DRNN}

Standard RNN architectures are inherently one dimensional. However, most interesting quantum many-body systems live in higher dimensions. By taking inspiration from Refs.~[\onlinecite{graves2012supervised}] and [\onlinecite{oord2016pixel}], we generalize one dimensional RNNs to multidimensional RNN wave functions. In particular, we generalize to 2D vanilla RNNs that are more suitable to simulating two-dimensional square lattices than one-dimensional RNNs, which map two-dimensional lattice configurations to one-dimensional configurations and do not necessarily encode spatial information about neighboring sites in a plausible manner. 

The main idea behind the implementation of 2D RNNs~\cite{graves2012supervised} is to replace the single hidden state that is passed from one site to another by two hidden states, with each one corresponding to the state of a neighboring site (vertical and horizontal) and hence respecting the 2D geometry of the problem. To do so, we change the one-dimensional recursion relation in \Eq{eq:rnn_action} to the two-dimensional recursion relation
\begin{equation}
    \bm{h}_{i,j} = f\!\left(W^{(h)} 
    [\bm{h}_{i-1,j} ; \bm{\sigma}_{i-1,j}] + W^{(v)} 
    [\bm{h}_{i,j-1} ; \bm{\sigma}_{i,j-1}] + \bm{b}\right),
    \label{eq:2drnn_action}
\end{equation}
where $\bm{h}_{i,j}$ is the hidden state at site $(i,j)$, $W^{(v,h)}$ are weight matrices and $\bm{b}$ is a bias. Here $f$ is a non-linear activation function chosen to be equal to the exponential linear unit (ELU) defined as
\begin{equation*}
    \text{ELU}(x) = \begin{cases}
		  x, & \text{if } x \geq 0\,,  \\
		\exp(x)-1, & \text{if } x < 0 \,.
	\end{cases}
\end{equation*}
The cost of computing a new hidden state $\bm{h}_{i,j}$ is quadratic in the size of the hidden state (number of memory units $d_h$), and the cost of computing the gradients with respect to the variational parameters of the 2D RNN remains unchanged. This property allows us to train 2D RNNs with a relatively large $d_h$.

To initialize the 2D RNN, we choose $\bm{h}_{i,0}, \bm{\sigma}_{i,0}$ and $\bm{h}_{0,j}, \bm{\sigma}_{0,j}$ to be null vectors. Once $\bm{h}_{i,j}$ is computed, we apply the same scheme as in Sec.~\ref{sec:RWF} to sample a spin $\sigma_{i,j}$. The scheme for computing positive or complex amplitudes from Sec.~\ref{sec:RWF} remains the same.

We note that generalization to higher dimensions, to other lattices, as well as to other types of RNN architectures can be done by taking inspiration from this scheme. For instance, using LSTMs~\cite{hochreiter1997long}, GRUs \cite{cho-etal-2014-properties} or Transformers \cite{Transformer} instead of vanilla RNNs in two dimensions is expected to make a significant improvement. We also expect that using multiplicative interactions \cite{wu2016multiplicative} might increase the expressiveness of 2D RNNs as compared to the additive interactions in \Eq{eq:2drnn_action}.

% ////////////////////////////////////////

\section{Variational Monte Carlo and Variance Reduction}
\label{sec:VMC}
The main goal of variational Monte Carlo (VMC) is to iteratively optimize an ansatz wave function to approximate, e.g., ground states of local Hamiltonians. 
VMC starts from a suitable {\it trial wave function} $\ket{\Psi_\lambda}$ that
incorporates the variational degrees of freedom of the approach.
$\ket{\Psi_\lambda}$ could be, for example, an MPS wave function \cite{sandvik_variational_2007}
in which case the free parameters are the MPS matrices.
Crucially, the ansatz $\ket{\Psi_\lambda}$ has to allow for {\it efficient sampling} from the square of the amplitudes
of $\ket{\Psi_\lambda}$. In this paper, we choose RNN wave functions, described in Sec.~\ref{sec:RWF}, to parametrize 
%a family of
the trial wave function $\ket{\Psi_\lambda}$ for a VMC optimization of ground states.

The aim of the VMC optimization is to minimize the expectation value of the
energy
\begin{align}
  E \equiv \frac{\braket{\Psi_\lambda|\hat{H}|\Psi_\lambda}}{\braket{\Psi_\lambda|\Psi_\lambda}}\label{eq:energy}
\end{align}
when given a family of states $\ket{\Psi_\lambda}$. 
This minimization is carried out using the gradient descent method or any of its variants.
Since the RNN wave function is normalized such that $\braket{\Psi_\lambda|\Psi_\lambda}=1$, the expectation value in
\Eq{eq:energy} can be written as
\begin{align}
  E = \braket{\Psi_\lambda |H|\Psi_\lambda}
  &=\sum_{\bm{\sigma}} |\psi_\lambda(\bm{\sigma})|^2\sum_{\bm{\sigma'}} H_{\bm{\sigma\sigma'}}\frac{\psi_\lambda(\bm{\sigma'})}{\psi_\lambda(\bm{\sigma})} \nonumber \\
  &\equiv \sum_{\bm{\sigma}} |\psi_\lambda(\bm{\sigma})|^2E_{loc}(\bm{\sigma}) \nonumber \\
   &\approx \frac{1}{N_S}\sum_{\bm{\sigma} \sim |\psi_\lambda(\bm{\sigma})|^2} E_{loc}(\bm{\sigma}),
  \label{eq:expectation_value} 
\end{align}
which represents a sample average of the local energy $E_{loc}(\bm{\sigma})$. The latter can be calculated efficiently for local Hamiltonians. Denoting $\lambda_i$ to be the real variational parameters of $\ket{\Psi_\lambda}$, the gradients $\partial_{\lambda_i}E$ can be similarly written as
\begin{align}
  \partial_{\lambda_i} E =
  \sum_{\bm{\sigma}} |\psi_\lambda(\bm{\sigma})|^2\frac{\partial_{\lambda_i}\psi^{*}_\lambda(\bm{\sigma})}{\psi^{*}_\lambda(\bm{\sigma})} E_{loc}(\bm{\sigma})  + \text{c.c}.
  \label{eq:gradient}
\end{align}
An optimization step consists of drawing $N_S$ samples $\{ \bm{\sigma^{(1)}}, \bm{\sigma^{(2)}}, \ldots, \bm{\sigma^{(N_S)}}\}$ from $|\psi_\lambda(\bm{\sigma})|^2$ autoregressively using the RNN wave function, and then
computing $\partial_{\lambda_i} E$ from \Eq{eq:gradient} as 
\begin{align}
\partial_{\lambda_i} E \approx
\frac{2}{N_S} \mathfrak{Re} \left ( \sum_{i=1}^{N_S}  \frac{\partial_{\lambda_i}\psi^{*}_\lambda(\bm{\sigma^{(i)}})}{\psi^{*}_\lambda(\bm{\sigma^{(i)}})} E_{loc}(\bm{\sigma^{(i)}}) \right),
\label{eq:stochastic_gradient1}
\end{align}
using automatic differentiation \cite{zhang2019automatic} and updating the parameters (if using gradient descent) according to
\begin{align}
  \lambda_i \leftarrow \lambda_i - \alpha \partial_{\lambda_i}E
  \label{eq:gradient_descent}
\end{align}
with a small learning rate $\alpha$. Instead of this simple gradient descent rule, we use the Adam optimizer \cite{kingma2014adam} to implement the gradient updates. We found that the latter gives better results compared to the simple gradient descent optimization shown in~\Eq{eq:gradient_descent} and without having to carefully tune the learning rate $\alpha$.

We note that the stochastic evaluation of the gradients in Eq.~\eqref{eq:stochastic_gradient1} tends to carry noise that increases their variances~\cite{Assaraf_1999,Clark2010}. Such high variances tend to slow down the convergence to the ground state energy. We propose to cure this limitation by introducing a new term in Eq.~\eqref{eq:stochastic_gradient1} that helps reduce the variance of the gradients by approximating
\begin{align}
\partial_{\lambda_i} E &\approx
\frac{2}{N_S} \mathfrak{Re} \left ( \sum_{i=1}^{N_S}  \frac{\partial_{\lambda_i}\psi^{*}_\lambda(\bm{\sigma^{(i)}})}{\psi^{*}_\lambda(\bm{\sigma^{(i)}})} \left ( E_{loc}(\bm{\sigma^{(i)}}) - E \right ) \right) \nonumber\\
 &=
\frac{2}{N_S} \mathfrak{Re} \left ( \sum_{i=1}^{N_S} \partial_{\lambda_i} \log \psi^{*}_\lambda(\bm{\sigma^{(i)}}) \left ( E_{loc}(\bm{\sigma^{(i)}}) - E \right ) \right),
\label{eq:stochastic_gradient2}
\end{align}
and we show below that this approximation does not introduce a bias.
This new term is useful for reducing the uncertainty in the gradient estimation, as in the limit where $E_{loc}(\bm{\sigma^{(i)}}) \approx E$ near convergence, the variance of the gradients $\partial_{\lambda_i} E$ goes to zero as opposed to the nonzero variance of the gradients in Eq.~\eqref{eq:stochastic_gradient1}. As a consequence, a stable convergence to the ground state is achieved as confirmed by our experiments. This idea is similar in spirit to control variate methods in Monte Carlo \cite{Assaraf_1999} and to baseline methods in reinforcement learning~\cite{mohamed2019monte}.

To show that the term we add in \Eq{eq:stochastic_gradient2} does not bias the true gradients in \Eq{eq:gradient}, it suffices to prove that
\begin{equation}
    \mathfrak{Re} \left(\left\langle \partial_{\lambda_i} \log \left( \psi^{*}_\lambda(\bm{\sigma}) \right) \right\rangle E \right) = 0,
    \label{eq:baseline2}
\end{equation}
where $\langle ... \rangle$ denotes the statistical average over the probability distribution $\lvert \psi_\lambda |^2$. 
To prove this expression, we write $\psi_\lambda(\bm{\sigma}) =  \sqrt{P_\lambda(\bm{\sigma})} \exp({\text i} \phi_\lambda(\bm{\sigma}))$, which implies that
\begin{equation*}
    \log \left(\psi^{*}_\lambda(\bm{\sigma}) \right) =  \frac12 \log \left( P_\lambda (\bm{\sigma}) \right) - {\text i} \phi_\lambda(\bm{\sigma}),
\end{equation*}
and hence
\begin{align}
     \mathfrak{Re} &\left (\left\langle \partial_{\lambda_i} \log \left( \psi^{*}_\lambda(\bm{\sigma}) \right) \right\rangle E \right) = \nonumber \\
     &\frac12 \left\langle \partial_{\lambda_i} \log \left( P_\lambda(\bm{\sigma}) \right) \right\rangle \mathfrak{Re}(E) + \left\langle \partial_{\lambda_i} \phi_\lambda(\bm{\sigma}) \right\rangle \mathfrak{Im}(E).
     \label{eq:baseline}
\end{align}
To show that $\left\langle \partial_{\lambda_i} \log \left( P_\lambda(\bm{\sigma}) \right) \right\rangle = 0$~\cite{sutton2000policy,mohamed2019monte}, we write
\begin{align*}
      \left\langle \partial_{\lambda_i} \log \left( P_\lambda(\bm{\sigma}) \right) \right\rangle &= \sum_{\bm{\sigma}} P_\lambda(\bm{\sigma}) ~ \partial_{\lambda_i} \log \left( P_\lambda(\bm{\sigma}) \right), \\
       &= \sum_{\bm{\sigma}} P_\lambda(\bm{\sigma}) \frac{\partial_{\lambda_i} P_\lambda(\bm{\sigma})}{P_\lambda(\bm{\sigma})},\\
       &= \partial_{\lambda_i} \sum_{\bm{\sigma}} P_\lambda(\bm{\sigma}), \\
       &= \partial_{\lambda_i} 1 = 0,
\end{align*}
where the fact that the RNN wave function is normalized justifies the  transition from the third line to the fourth line. 

From here, it suffices to show that $\left\langle \partial_{\lambda_i} \phi_\lambda(\bm{\sigma}) \right\rangle \mathfrak{Im}(E) = 0$. Since the Hamiltonian $\hat{H}$ is Hermitian, the expectation value $E$ is real and hence $\mathfrak{Im}(E) = 0$.
We therefore arrive at \Eq{eq:baseline2}.

% ///////////////////////////////////////////
\section{Implementing Symmetries}
\label{sec:symmetries}
\subsection{Imposing discrete symmetries}
\label{sec:discrete_symmetries}

Inspired by Refs.~[\onlinecite{Wu_2019}] and [\onlinecite{sharir2019deep}], we propose to implement discrete symmetries in a similar fashion for RNN wave functions without spoiling their autoregressive nature.

Assuming that a Hamiltonian $\hat{H}$ has a symmetry under discrete transformations $\mathcal{T}$, its ground state
\begin{equation*}
    \ket{\Psi_G} = \sum_{\bm{\sigma}}\psi_G(\bm{\sigma})\ket{\bm{\sigma}}
\end{equation*}
is an eigenvector of the symmetry transformation $\mathcal{T}$. The ground state transforms as $\psi_G(\mathcal{T}\bm{\sigma}) = \omega_\mathcal{T} \psi_G(\bm{\sigma})$ where $\omega_\mathcal{T}$ is an eigenvalue with module 1, that is independent of the choice of $\bm{\sigma}$. This expression implies that the transformation $\mathcal{T}$ changes the ground state with only a global phase term that does not affect the probability distribution, and changes the sign structure with a global phase term. 
It is thus desirable that the RNN wave function also has this symmetry.

To enforce a discrete symmetry $\{\mathcal{T}\}$ on an RNN wave function $\ket{\Psi_\lambda}$, we propose the following scheme:
\begin{itemize}
    \item Generate a sample $\bm{\sigma}$ autoregressively from the RNN wave function.
    \item Sample with a probability $1/\text{Card}(G)$ a transformation $\mathcal{T}$ from the symmetry transformation group $G = \{\mathds{1},\mathcal{T}_1,...\}$ that leaves the Hamiltonian $\hat{H}$ invariant, and apply the transformation $\mathcal{T}$ to $\bm{\sigma}$.
    \item Assign to the spin configuration $\bm{\tilde{\sigma}} = \mathcal{T}\bm{\sigma}$ the amplitude $\psi_\lambda(\bm{\tilde{\sigma}}) = \sqrt{P_\lambda(\bm{\tilde{\sigma}})} \exp(i \phi_\lambda(\bm{\tilde{\sigma}}))$, such that
    \begin{align*}
        P_\lambda(\bm{\tilde{\sigma}}) &= \frac{1}{\text{Card(G)}} \left ( \sum_{\mathcal{\tilde{T}} \in G} P_\lambda\left(\mathcal{\tilde{T}}\bm{\sigma}\right) \right ), \\
        \phi_\lambda(\bm{\tilde{\sigma}}) &= \text{Arg} \left (\omega_\mathcal{T} \sum_{\mathcal{\tilde{T}} \in G} \exp\left(\text{i} \phi_\lambda\left(\mathcal{\tilde{T}}\bm{\sigma}\right)\right) \right ) ,
    \end{align*}
    where $P_\lambda(\mathcal{\tilde{T}}\bm{\sigma})$ is a probability generated by the Softmax layer and $\phi_\lambda(\mathcal{\tilde{T}}\bm{\sigma})$ is a phase generated by the Softsign layer, as explained in Sec. \ref{sec:RWF}. 
\end{itemize}
If the ground state is positive~\cite{Bravyi:2008:CSL:2011772.2011773}, we use the same algorithm but only symmetrize the probability $P_\lambda$, without having to worry about symmetrizing the phase $\phi_\lambda$. 

For concreteness, we illustrate the algorithm above with ``Symmetric RNNs'' that have a built-in parity symmetry. We use this architecture in Sec.~\ref{sec:1DTFIMresults} to get a more accurate estimate of the ground state of the 1D TFIM that also obeys a parity symmetry. Indeed, symmetric RNNs show an improvement over ordinary pRNN wave functions on the task of estimating the second Rényi entropy (see App.~\ref{sec:entropy}). Symmetric RNNs can be implemented using the following procedure: 
\begin{itemize}
    \item Sample each configuration $\bm{\sigma}$.
    \item Choose to apply or to not apply the parity transformation $\hat{\mathcal{P}}$ on $\bm{\sigma}$ with a probability $1/2$.
    \item Assign to $\bm{\sigma}$ the probability: 
    \begin{equation*}
        P = \frac{\left(P_{\lambda}\left(\bm{\sigma}\right)+P_{\lambda}\left(\hat{\mathcal{P}}\bm{\sigma}\right)\right)}{2}.
    \end{equation*}
\end{itemize}
We also emphasize the possibility of carefully designing RNN wave functions to impose discrete symmetries, without using the symmetrization scheme above and which we leave for future investigations.

\subsection{Imposing zero magnetization}
\label{sec:zeromag}

Since the ground state of the $J_1$-$J_2$ model has zero magnetization, i.e., a $U(1)$ symmetry~\cite{Marshall1955, Lieb1962}, it is helpful to enforce this constraint on our RNN wave functions to get accurate estimations of the ground state energy.
To do so, we propose an efficient way to generate samples with zero magnetization while maintaining the autoregressive property of the RNN wave function. The procedure effectively applies a projector  $\mathcal{P}_{S_z=0}$ to the original state, which restricts the RNN wave function
to the subspace of configurations with zero magnetization. This procedure avoids generating a large number of samples and discarding the ones that have non-zero magnetization. 

The condition of zero magnetization implies that the number of up spins should be equal to the number of down spins. To satisfy this constraint, we utilize the following algorithm:
\begin{itemize}
    \item Sample autoregressively the first half of the spin configuration $(\sigma_1,\sigma_2,..., \sigma_{N/2})$
    \item At each step $i>N/2$:
    \begin{itemize}
        \item Generate the output of the RNN wave function: $\bm{y}_i = (\psi_i^{\text{down}}, \psi_i^{\text{up}})$ where $\psi_i^{\text{down}}$ and $\psi_i^{\text{up}}$ are both non-zero and their modules squared sum to $1$.
        \item Define the following amplitudes:
        \begin{align*}
            a_i &=  \psi_i^{\text{down}} \times \Xi\left(\frac{N}{2} - N_{\text{down}}(i)\right), \\
            b_i &=  \psi_i^{\text{up}} \times \Xi\left(\frac{N}{2} - N_{\text{up}}(i)\right),
        \end{align*} 
        where
        \begin{equation*}
                \Xi(x) \equiv \begin{cases}
                		  1, & \text{if } x > 0\,,  \\
                		0, & \text{if } x \leq 0 \,,
                \end{cases}
        \end{equation*}
        and 
        \begin{align*}
            \qquad \qquad N_{\text{down}}(i) &= \text{Card}\left(\{ j \ / \sigma_j = 0 \  \text{and} \ j<i \}\right) , \\ 
            \qquad \qquad N_{\text{up}}(i) \quad &= \text{Card}\left(\{  j \ /  \sigma_j = 1 \  \text{and} \ j<i \}\right).
        \end{align*}
        In words, $N_{\text{up}}(i)/N_{\text{down}}(i)$ is the number of up/down spins generated before step $i$.
        \item Sample $\sigma_i$ from $|\bm{\tilde{y}}_i|^2$, where:
        \begin{equation*}
            \bm{\tilde{y}}_i = \frac{1}{{\sqrt{a_i^2+b_i^2}}}(a_i,b_i)
        \end{equation*}
        which is normalized, i.e. $\lvert \lvert \bm{\tilde{y}}_i\rvert \rvert_2 = 1$.
    \end{itemize}
\end{itemize}
Using this algorithm, it is clear that the RNN wave function generates a spin configuration that has the same number of up spins and down spins, and hence a zero magnetization. In fact, at each step $i>N/2$, the function $\Xi$ assigns a zero amplitude for the next spin $\sigma_i$ to be spin up if $N_{\text{up}}(i) = N/2$ or to be spin down if $N_{\text{down}}(i) = N/2$. 

Interestingly enough, our scheme does not spoil the normalization of the RNN wave function as the new conditional probabilities $|\bm{\tilde{y}}_i|^2$ are also normalized. We also note that this algorithm preserves the autoregressive property of the original RNN wave function and can also be parallelized. Moreover, this scheme can be easily extended to the generation of samples with a non-zero fixed magnetization, which is useful when considering the problem of finding states that live in a non-zero fixed magnetization sector.

% ///////////////////////////////////////////
\section{R\'enyi entropies}
\label{sec:entropy}
Given a quantum system with a spatial bipartition $(A,B)$, one can write the RNN wave function $\ket{\Psi_\lambda}$ as
\begin{equation*}
    \ket{\Psi_\lambda} = \sum_{\bm{\sigma}_{A}, \bm{\sigma}_{B}}  \psi_\lambda(\bm{\sigma}_{A}\bm{\sigma}_{B}) \, \ket{\bm{\sigma}_{A} \bm{\sigma}_{B}},
\end{equation*}
where $\bm{\sigma}_{A/B}$ denotes the spin configuration that lives in the partition $A/B$ and $\bm{\sigma}_A \bm{\sigma}_B$ stands for a concatenation of $\bm{\sigma}_A$ and $\bm{\sigma}_B$.

\begin{figure}
    \centering
    \includegraphics[width = 0.9\linewidth]{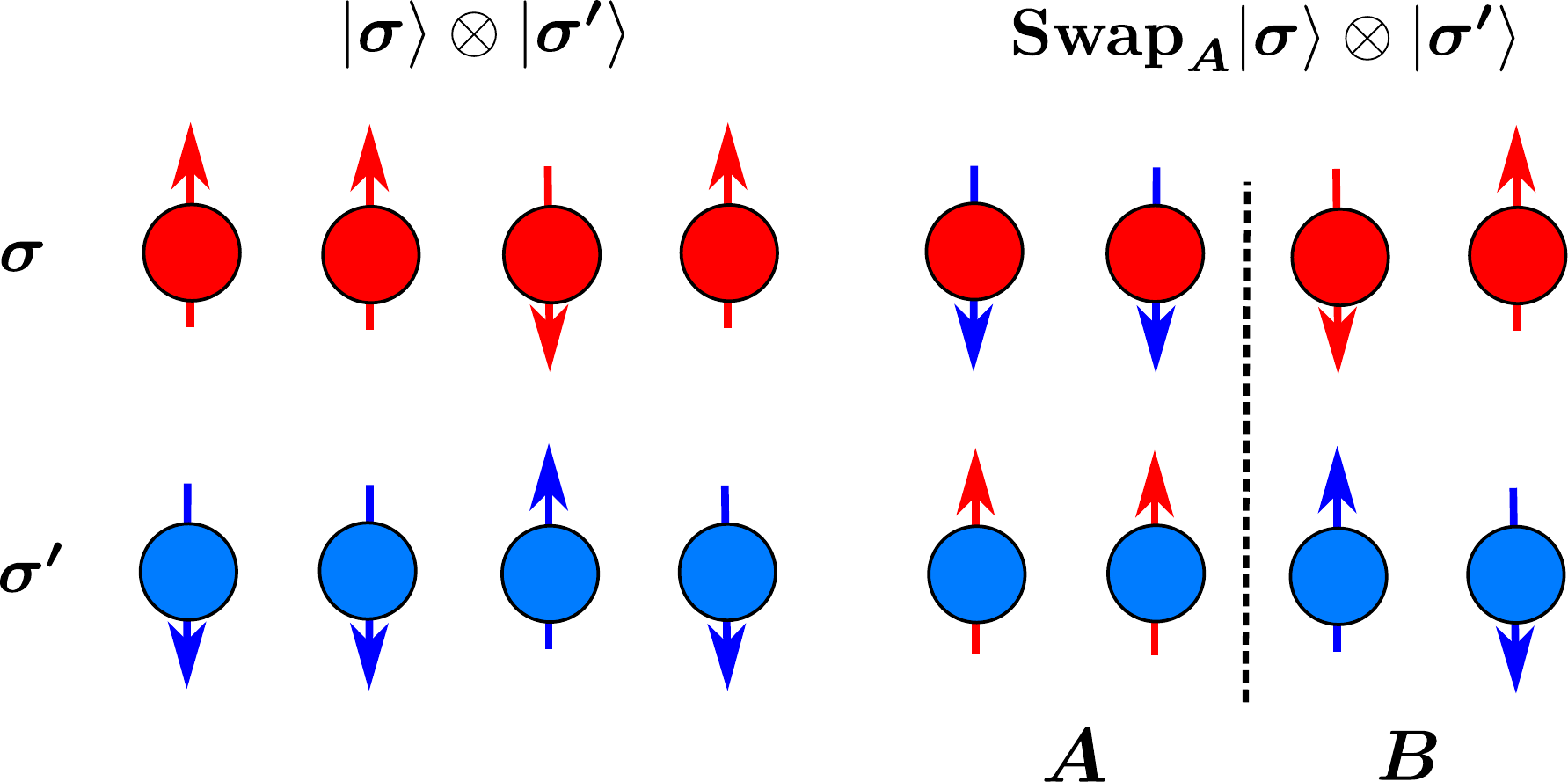}
    \caption{The Swap operator acting on the tensor product of two samples \bm{$\sigma$} and \bm{$\sigma'$}.}
    \label{fig:Replica}
\end{figure}

The $\alpha$-R\'enyi entropy between region $A$ and $B$ is given by
\begin{equation}
    S_\alpha (A) = \frac{1}{1-\alpha} \log \left ( \text{Tr} \rho_A^\alpha \right ),
\end{equation}
where $\rho_A = \text{Tr}_B \ket{\Psi_\lambda}\bra{\Psi_\lambda}$ and $\alpha$ is an integer~\cite{hastings2010measuring}. To estimate these entropies, we use the so-called replica trick~\cite{hastings2010measuring}, where we consider the action of the $\text{Swap}_A$ operator on the two copies of the RNN wave function, which swaps the spins in the region $A$ between the two copies (as demonstrated in Fig.~\ref{fig:Replica}) such that
\begin{align}
  &\text{Swap}_A \ket{\Psi_\lambda} \otimes \ket{\Psi_\lambda} \nonumber \\
  &= \sum_{\bm{\sigma}, \, \bm{\tilde{\sigma}}}
     \psi_\lambda(\bm{\sigma}_A \bm{\sigma}_B) \psi_\lambda(\bm{\tilde{\sigma}}_A \bm{\tilde{\sigma}}_B) \ket{\bm{\tilde{\sigma}}_A \bm{\sigma}_B} \otimes \ket{\bm{\sigma}_A \bm{\tilde{\sigma}}_B}.
\end{align}
The expectation value of $\text{Swap}_A$ in the double copy of the RNN wave function ``$\ket{\Psi_\lambda} \otimes \ket{\Psi_\lambda}$'' is given by \cite{hastings2010measuring,torlai2018}
\begin{align}
  \langle \text{Swap}_A \rangle &= \sum_{\bm{\sigma}, \, \bm{\tilde{\sigma}}}
     \psi_\lambda^{*}(\bm{\sigma}_A \bm{\sigma}_B) \psi_\lambda^{*}(\bm{\tilde{\sigma}}_A \bm{\tilde{\sigma}}_B) \psi_\lambda(\bm{\tilde{\sigma}}_A \bm{\sigma}_B) \psi_\lambda(\bm{\sigma}_A \bm{\tilde{\sigma}}_B)
     \nonumber \\
     &= \text{Tr} \rho^2_A = \exp (-S_2(A)).
     \label{eq:exp_renyi}
\end{align}
Hence, by calculating the expectation of the value of the Swap operator in the double copy of the RNN wave function, we can access the second R\'enyi entropy. Interestingly, the R\'enyi entropies $S_\alpha$ have been shown to encode similar properties independently of $\alpha$~\cite{hastings2010measuring, Alioscia2009}. 

Although an exact evaluation of Eq.~\eqref{eq:exp_renyi} is numerically intractable, we can use importance sampling to estimate it \cite{hastings2010measuring} as
\begin{align}
    &\langle \text{Swap}_A \rangle 
    \nonumber \\
    &= \sum_{\bm{\sigma}, \, \bm{\tilde{\sigma}}}
     |\psi_\lambda(\bm{\sigma}_A \bm{\sigma}_B)|^2 | \psi_\lambda
     (\bm{\tilde{\sigma}}_A \bm{\tilde{\sigma}}_B)|^2 \frac{\psi_\lambda(\bm{\tilde{\sigma}}_A \bm{\sigma}_B) \psi_\lambda(\bm{\sigma}_A \bm{\tilde{\sigma}}_B)}{\psi_\lambda(\bm{\sigma}_A \bm{\sigma}_B) \psi_\lambda
     (\bm{\tilde{\sigma}}_A \bm{\tilde{\sigma}}_B)}, \nonumber
     \\
     &\approx \frac{1}{N_s} \sum_{i=1}^{N_s} \frac{\psi_\lambda(\bm{\tilde{\sigma}^{(i)}}_A \bm{\sigma^{(i)}}_B) \psi_\lambda(\bm{\sigma^{(i)}}_A \bm{\tilde{\sigma}^{(i)}}_B)}{\psi_\lambda(\bm{\sigma^{(i)}}_A \bm{\sigma^{(i)}}_B) \psi_\lambda
     (\bm{\tilde{\sigma}^{(i)}}_A \bm{\tilde{\sigma}^{(i)}}_B)}.
\end{align}
Using this trick, for the system sizes studied in this paper we only have to generate two sets of exact samples $\{\bm{\sigma^{(i)}}\}_{i=1}^{N_s}$ and  $\{\bm{\tilde{\sigma}^{(i)}}\}_{i=1}^{N_s}$ independently from $\lvert \psi_\lambda \rvert^2$ without having to use the improved ratio trick~\cite{hastings2010measuring}. By defining
\begin{equation*}
     \text{Swap}^{(i)}_A \equiv \frac{\psi_\lambda(\bm{\tilde{\sigma}^{(i)}}_A \bm{\sigma^{(i)}}_B) \psi_\lambda(\bm{\sigma^{(i)}}_A \bm{\tilde{\sigma}^{(i)}}_B)}{\psi_\lambda(\bm{\sigma^{(i)}}_A \bm{\sigma^{(i)}}_B) \psi_\lambda
     (\bm{\tilde{\sigma}^{(i)}}_A \bm{\tilde{\sigma}^{(i)}}_B)},
\end{equation*}
the statistical error on the estimation of the R\'enyi-2 entropy can be calculated as
\begin{equation*}
    \epsilon = \frac{1}{\langle \text{Swap}_A \rangle } \sqrt{\frac{\text{var}\left(\{\text{Swap}^{(i)}_A\} \right)}{N_s}}.
\end{equation*}

For the estimation of the R\'enyi-2 entropy for the 1D TFIM in this paper, we use $N_s = 2 \times 10^6$ samples from a trained RNN wave function with one GRU layer and $50$ memory units.

During the completion of this paper, we became aware of another way to estimate entanglement entropies using autoregressive models with conditional sampling~\cite{NAQSentropy}.

% /////////////////////////////////////////////
\section{Tables of Results}
\label{sec:tables_of_results}

In Tab.~\ref{tab:cRNNvsDMRG}, we state the variational energies of the cRNN wave function for the 1D $J_1$-$J_2$ model and compare with results from DMRG. 
We examine two different methods of training. 
First, we do not impose an initial sign structure while, secondly, we introduce a background Marshall sign. 
The results suggest that using a Marshall sign improves the results significantly for $J_2 = 0.0,0.2$ and $0.5$ (with $J_1 =1$ for all cases).
\begin{table}[bt]
\setlength\extrarowheight{2pt}
\centering
\begin{tabular}{|c|c|c|c|}\hline
\multirow{2}{*}{$J_2$} &  \multicolumn{3}{c|}{$E/N$} \\
\cline{2-4}
& No Sign & Marshall Sign & DMRG \\
%\hhline{|=|===|}
\hline
$0.0$ & -0.4412480(2) & -0.4412760(1) & -0.4412773 \\ 
$0.2$ & -0.4073635(3) & -0.4073871(3) & -0.4073881 \\ 
$0.5$  & -0.3749958(6) & -0.3750006(6) & -0.3750000  \\ 
$0.8$  & -0.4205478(13) & -0.4205695(12) & -0.4207006  \\ \hline
\end{tabular}
\caption{Energy per spin values for the 1D $J_1$-$J_2$ model.
We consider a cRNN wave function with two different methods of training (with no initial sign structure and with a background Marshall sign) and compare with results from DMRG. 
All results correspond to $100$ spins and have $J_1=1$.
We use three GRU layers, where each layer has $100$ units.
Note that $J_2=0.5$ corresponds to the Majumdar-Ghosh model where the ground state is a product-state of spin singlets. For the estimation of the variational energies we use $4 \times 10^6$ samples.}
\label{tab:cRNNvsDMRG}
\end{table}

In Tab.~\ref{tab:2DComparison}, we compare the variational energies per site of the 2D TFIM with a lattice size of $12 \times 12$ for different values of the transverse magnetic field $h$, for a 1D pRNN wave function, a 2D pRNN wave function, a PixelCNN wave function \cite{sharir2019deep} and DMRG.

\begin{table*}[bt]
\setlength\extrarowheight{2pt}
\centering
\begin{tabular}{|c|c|c|c|c|}\hline
\multirow{2}{*}{$h$} &  \multicolumn{4}{c|}{$E/N$} \\
\cline{2-5}
& 1DRNN & 2DRNN  & PixelCNN & DMRG \\ \hline
$2$  & -2.4096018(2) & \textbf{-2.40960262(9)} & -2.4096022(2) & \textbf{-2.40960263} \\ %\hline
$3$  & -3.1738969(5) & \textbf{-3.1739018(2)} & -3.1739005(5) & -3.17389966 \\ %\hline
$4$  & -4.1217969(3) & \textbf{-4.12179808(6)} & -4.1217979(2) & -4.12179793  \\ \hline
\end{tabular}
\caption{Variational energies per site for a 1D pRNN wave function (three layers of GRUs with 100 memory units), 2D pRNN wave function (a single layer of 2D vanilla RNN with 100 memory units), PixelCNN wave functions with results taken from Ref.~[\onlinecite{sharir2019deep}] and DMRG (with bond dimension $\chi = 512$ for $h = 2$ and $\chi = 1024$ for both $h = 3,4$). As a benchmark, we use the 2D TFIM with a lattice size of $12 \times 12$ for different values of $h$ where the critical point is at $h \approx 3$. Values in bold font correspond to the lowest variational energies and hence to the most accurate estimations of the ground state energy across all four methods. For the estimation of the variational energy of the trained 1D and 2D pRNN wave functions, we use $2 \times 10^6$ samples.}
\label{tab:2DComparison}
\end{table*}

% ////////////////////////////////////////////
\section{Scaling of resources (continued)}
\label{sec:scaling_app}

\begin{figure}[t]
    \centering
    \includegraphics[width = \columnwidth]{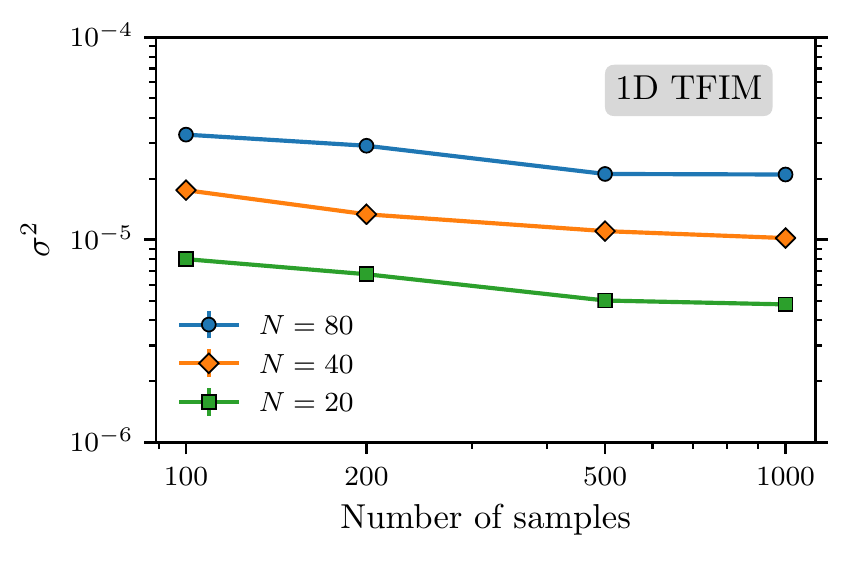}
    \caption{The energy variance per spin against the number of samples, which suggests that the energy variance saturates and does not improve further by using a larger number of samples for training.}
    \label{fig:Scaling_study_2}
\end{figure}

\Fig{fig:Scaling_study_2} shows the dependence of $\sigma^2$ on the number of samples used to estimate the gradients of the variational energy (see App.~\ref{sec:VMC}). We investigate this effect for the case of the 1D TFIM, using 50 memory units in the pRNN wave function.
Even though a large number of samples yields higher statistical accuracy of the gradient estimates used in our optimizations, we observe only a weak dependence of $\sigma^2$ on the number of samples for all studied system sizes.

\begin{figure}[t]
    \centering
    \includegraphics[width = \columnwidth]{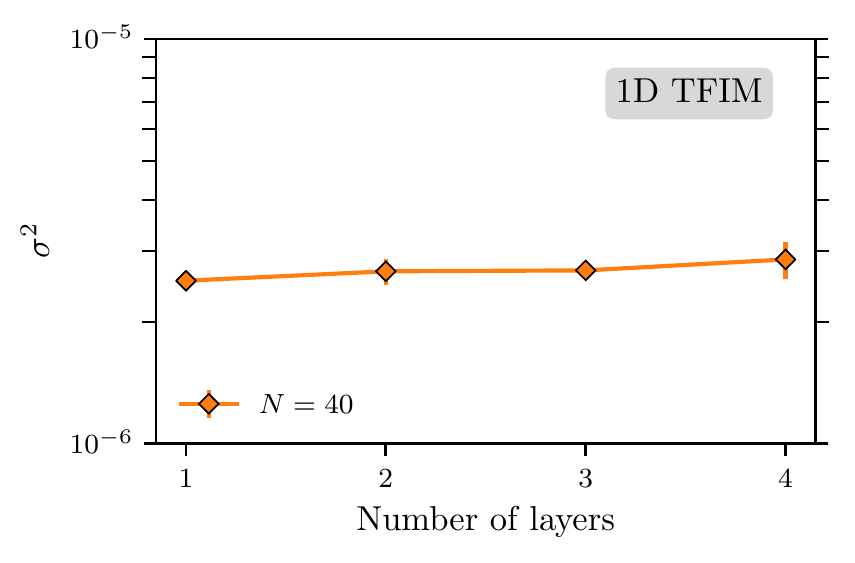}
    \caption{Scaling study of the energy variance per spin vs the number of layers of a pRNN wave function such that all pRNN wave functions with different layers have approximately the same number of variational parameters. The results show that fixing the number of parameters while changing the number of layers does not affect the energy variance obtained by the pRNN wave function.}
    \label{fig:Scaling_study_3}
\end{figure}

In \Fig{fig:Scaling_study_3} we present results for the dependence of $\sigma^2$
on the depth of the pRNN wave function architecture for a critical TFIM with $N=40$ sites. We 
investigate architectures up to a depth of four layers. The number of memory units is adapted such that we have a similar number of variational parameters ($\sim$31000) for each of the four architectures. We find that $\sigma^2$ depends only weakly on the number
of layers. 

% /////////////////////////////////////////////
% \newpage
\section{Hyperparameters}
\label{sec:hyperparams}
In Tab.~\ref{tab:hyperparams}, we present the hyperparameters used to train the RNN wave functions in this paper. 
We anticipate that further improvements such as the use of stochastic reconfiguration \cite{becca2017} or a computationally cheaper variant such as K-FAC \cite{martens2018kronecker} for the optimization could potentially lead to more accurate estimations of the ground state energies as compared to the Adam optimizer~\cite{kingma2014adam}.
Seeds are listed in the table for reproducibility purposes.

\begin{table*}[p]
    \centering
    \begin{tabular}{|c|c|c|}\hline
       Figures & Hyperparameter & Value \\\hline
        \multirow{5}{*}{Fig. \ref{fig:1d_TFIM_results}} & Architecture & One-layer 1D pRNN wave function with $50$ memory units\\
            & Number of samples & $N_s = 1000$ ($N=20$), $N_s = 500$ ($N=80$), $N_s = 200$ ($N=1000$) \\
             & Training iterations & $20000$  \\
             & Learning rate & $5 \times 10^{-3}$\! \! \\
             & Seed & $111$  \\
       \hline
      \multirow{5}{*}{Fig. \ref{fig:J1J2_results}} & Architecture & Three-layer 1D cRNN wave function with $100$ memory units\\
      & Number of samples & 500 \\
      &  Training iterations & $100000$ \\
       & Learning rate & $(\eta^{-1} + 0.1t)^{-1}$ with $\eta = 2.5 \times 10^{-4}$ \\
        &    Seed & $111$ \\
       \hline
             
       \multirow{5}{*}{Fig. \ref{fig:2DTFIM_study}(c): 1DRNN} & Architecture & Three-layer 1D pRNN wave function with $100$ memory units \\
       & Number of samples & $500$ \\
              &  Training iterations & $150000$ \\
              &  Learning rate & $(\eta^{-1} + 0.1t)^{-1}$ with $\eta = 10^{-3}$ \\
              &  Seed & $333$ \\
       \hline
     \multirow{5}{*}{Fig. \ref{fig:2DTFIM_study}(c): 2DRNN} & Architecture & One-layer 2D pRNN wave function with $100$ memory units \\
     & Number of samples & $500$ \\
              &  Training iterations & $150000$ \\
              &  Learning rate & $\eta(1 + t/5000)^{-1}$ with $\eta = 5 \times 10^{-3}$ \\
              &  Seed & $111$ \\
       \hline
       \multirow{5}{*}{Fig. \ref{fig:Scaling_study}(a)} & Architecture & One-layer 1D pRNN wave function \\
       & Number of samples & $500$ \\
       & Training iterations & $10000$ \\
       & Learning rate & $10^{-3}$ \\
      & Seeds & $111,222,333,444,555$ \\
        \hline
         \multirow{5}{*}{Fig. \ref{fig:Scaling_study}(b)} & Architecture & One-layer 1D pRNN wave function \\
         & Number of samples & $500$ \\
        & Training iterations & $10000$ \\
        & Learning rate & $(\eta^{-1} + 0.1t)^{-1}$ with $\eta = 10^{-3}$ \\
        & Seeds & $111,222,333,444,555,666,777,888,999,1111$ \\
       \hline
       \multirow{4}{*}{Fig. \ref{fig:Scaling_study_2}} & Architecture & One-layer 1D pRNN wave function with $50$ memory units \\
        & Training iterations & $10000$ \\
        & Learning rate & $10^{-3}$ \\
        & Seeds & $111,222,333,444,555$ \\
       \hline
        \multirow{5}{*}{Fig. \ref{fig:Scaling_study_3}} & Architecture & 1D pRNN wave function \\
        & Number of samples & $500$ \\
        & Training iterations & $10000$ \\
        & Learning rate & $5\times 10^{-3}$ \\
        & Seeds & $111,222,333,444,555$ \\
       \hline
       
    \end{tabular}
    \caption{Hyperparameters used to obtain the results reported in this paper. Note that the number of samples stands for the batch size used to train the RNN wave function. Multiple seeds are used for the scaling of resources study to provide error bars on our results.}
    \label{tab:hyperparams}
\end{table*}

%%%%%%%%%%%%%%%%%% REFERENCES %%%%%%%%%%%%%%%%%%
\clearpage

\end{document}